%% file: paper.tex
\documentclass{tlp}

\newtheorem{definition}{Definition} 
\newtheorem{example}{Example} 
\newtheorem{theorem}{Theorem}

\newcounter{ex:p:2}
\newcounter{ex:pq}
\newcounter{div ill}
\newcounter{ex:pq:graph}
\newcounter{dmd}
\newcounter{problem ex}
\newcounter{ex:pp:version1}

\usepackage{amssymb}
\usepackage{amsmath}
\usepackage{latexsym}
\usepackage{ifthen}
\usepackage[usenames]{color}
\usepackage{longtable}
\usepackage[latin1]{inputenc}
\usepackage{pdfpages}
\usepackage[hyperfootnotes=false,bookmarksopen=true]{hyperref}
\usepackage{mathpartir}
\usepackage{tikz}
\usetikzlibrary{arrows,decorations.pathmorphing,backgrounds,fit,positioning}
\usepackage{wrapfig}
\usepackage{setspace}

\input{commands.tex}

\input xy
\xyoption{all}

\renewcommand{\theequation}{\arabic{equation}}

\begin{document}
\bibliographystyle{acmtrans}

\title[Automated Termination Analysis for Logic Programs with
Cut]{Automated Termination Analysis for Logic Programs with
Cut\thanks{Supported by the Deutsche Forschungsgemeinschaft (DFG) under grant GI 274/5-2, the DFG Research Training Group 1298 (\emph{AlgoSyn}),
and the Danish Natural Science Research Council.}}
\author[P.\ Schneider-Kamp et al.]{PETER SCHNEIDER-KAMP \\
Dept.\ of Mathematics and Computer Science,
University of Southern Denmark,
Denmark
\and
J\"URGEN GIESL, THOMAS STR\"ODER \\
LuFG Informatik 2,
RWTH Aachen University,
Germany
\and
ALEXANDER SEREBRENIK \\
Dept.\ of Mathematics and Computer Science, TU Eindhoven,
The Netherlands
\and
REN\'E THIEMANN\\
Institute of Computer Science, 
University of Innsbruck,
Austria
}

\maketitle

\label{firstpage}

\begin{abstract}
Termination
is an important and well-studied property for logic programs. However, almost all
approaches for automated termination analysis focus on definite logic programs, whereas real-world \textsf{Prolog} programs typically use the \emph{cut} operator.
We introduce a novel
pre-processing method which automatically
transforms \textsf{Prolog} programs into logic programs
without cuts, where termination of the
cut-free program implies termination of the original program. Hence
after this pre-processing, any technique for proving termination of definite
logic programs can be applied.
We implemented this pre-processing in our termination prover \aprove\
and evaluated it successfully with extensive experiments.
\end{abstract}
%
%
\section{Introduction}\label{sec:introduction}

Automated termination analysis for logic programs has been widely studied,
see,\linebreak e.g.,
(\citeNP{CodishTOPLAS}; \citeNP{Codish:Lagoon:Stuckey};
\citeNP{DeSchreye:Decorte:NeverEndingStory};\linebreak  \citeNP{Mesnard:Serebrenik};
\citeNP{TPLP10}; \citeNP{TOCL08};\linebreak \citeNP{SerebrenikS05}). Still, virtually all existing techniques only prove
universal termination of \emph{definite} logic
programs, which do not use the \emph{cut} ``!''.
An exception
is  \cite{Marchiori:AMAST96}, which  transforms
``safely typed'' logic
programs to term rewrite systems (TRSs). However, the resulting
TRSs are complex and since there is no implementation
of
\cite{Marchiori:AMAST96}, it is unclear whether they
can be handled by existing TRS termination tools. Moreover, \pagebreak
\cite{Marchiori:AMAST96}'s method does not allow arbitrary cuts (e.g.,
it does not operate on programs like  \rEX{ex:divminus}).

In the present paper, we introduce a novel approach which shows that
universal termination
of logic programs with cuts can indeed be proved \emph{automatically}
for  (typically infinite)
classes of queries.
This  solves an important open problem in automated
termination analysis of logic programs.

\begin{example}
\label{ex:divminus}
We want to prove termination of the following program for the class of queries {\small $\{\Fdiv(t_1,t_2,t_3)\!\mid\!t_1,t_2\text{ are ground}\}$}.
Since we only regard programs without pre-defined
predicates, the program 
contains clauses defining
predicates for failure and equality. So the atom $\Ffailure(\Fa)$ always fails
and corresponds to \textsf{Prolog}'s pre-defined ``\textsf{fail}''.

\vspace*{-.25cm}

\hspace*{-.9cm}
{\footnotesize
\begin{minipage}[t]{200pt}{
\begin{eqnarray}
\label{ex:divminus:1}{\sf div}(X, \Fz, Z) & \hspace*{-.2cm}\from\hspace*{-.2cm} & !, \Ffailure(\Fa).\\
\label{ex:divminus:2}{\sf div}(\Fz, Y, Z) & \hspace*{-.2cm}\from\hspace*{-.2cm} & !, {\sf eq}(Z,\Fz).\\
\label{ex:divminus:3}{\sf div}(X, Y, \Fs(Z)) & \hspace*{-.2cm}\from\hspace*{-.2cm} & \Fminus(X,Y,U), {\sf
div}(U,Y,Z).\\
\label{ex:fail}\Ffailure(\Fb). & &
\end{eqnarray}}
\end{minipage}\begin{minipage}[t]{170pt}{
\begin{eqnarray}
\label{ex:divminus:7}{\sf eq}(X, X). & &\\
\label{ex:divminus:4}\Fminus(\Fz, Y, \Fz). & &\\
\label{ex:divminus:5}\Fminus(X, \Fz, X). & &\\
\label{ex:divminus:6}\Fminus(\Fs(X), \Fs(Y), Z) & \hspace*{-.2cm}\from\hspace*{-.2cm} & \Fminus(X,Y,Z).
\end{eqnarray}}
\end{minipage}}

\vspace*{-.15cm}

\noindent
Any 
termination analyzer that ignores the cut fails,
as $\Fdiv(\Fz,\Fz,Z)$ would lead to the subtraction of $\Fz$ and
start an infinite derivation using Clause (\ref{ex:divminus:3}).
So due to the cut, (universal) termination effectively depends on the order of the clauses.
\end{example}

There are already several static analysis techniques for logic
programming with cut, e.g.,
\cite{DBLP:conf/lpar/FileR93,DBLP:conf/ershov/Mogensen96}, which are based on
abstract interpretation \cite{DBLP:journals/jlp/CousotC92,%
DBLP:conf/slp/CharlierRH94,DBLP:conf/amast/SpotoL98}.
However,  these works do not capture termination as an observable and
none of these
results targets termination analysis explicitly.
While
we also rely on the idea of
abstraction, our approach does
not operate directly on the abstraction. Instead, we synthesize a cut-free
logic program
from the abstraction, such that termination of the derived program implies termination of
the original one. Thus, we can  benefit from the large body of existing work on
termination analysis for cut-free programs. Our
approach is inspired by our previous successful technique for termination
analysis of \textsf{Haskell} programs \cite{RTA06}, which in turn was inspired
by related approaches to program optimization \cite{SG95}.

In \rSC{sec:pet:concrete},  we introduce the required
notions and present a set of simple inference rules that characterize
logic programming with cut for concrete queries. In  \rSC{sec:pet:abstract} we extend these inference
rules
 to handle \emph{classes} of queries. Using these rules
we can automatically build so-called termination graphs, cf.\
\rSC{sec:pet:termgraph}.  Then, \rSC{sec:pet:graphs} shows how
to generate a new cut-free logic program from such a graph automatically.

Of course, one can transform any Turing-complete
formalism like logic programming with cuts into another Turing-complete
formalism like cut-free logic programming.
But the challenge is to develop a transformation such that termination of
the resulting
programs  is \emph{easy to analyze by existing termination
tools}. Our implementation and extensive experiments in
\rSC{sec:pet:conclusion} show that with our approach,
the resulting cut-free program is usually easy to handle by
existing  tools.


\section{Concrete Derivations}\label{sec:pet:concrete}

See e.g.\
\cite{Apt:Book} for  the basics of logic
programming.
We \pagebreak distinguish between individual cuts to make their scope
explicit. So a  signature $\Sigma$ contains   all
predicate and function symbols and all labeled versions of the cut
 $\{!_{m}/0 \mid m \in \N\}$. 
For simplicity we just consider terms
$\TT(\Sigma,\VV)$ and no atoms, i.e.,
we do not distinguish between predicate and function
symbols. To ease the presentation, in the paper
we exclude terms with cuts $!_m$ as proper subterms.
 A \emph{clause} is a pair $H \from B$ where
the \emph{head} $H$ is from $\TT(\Sigma,\VV)$ and  the
\emph{body} $B$ is a sequence of terms from $\TT(\Sigma,\VV)$.
Let $\Goal(\Sigma,\VV)$ be the set of all such sequences, where
$\Box$  is the empty goal.

A \emph{program} $\PP$ (possibly with cut) is a finite sequence of
clauses.
$\Slice(\P,t)$ are
all clauses for $t$'s predicate, i.e., {\small $\Slice(\P,p(t_{1},...,t_{n}))
= \{c \mid c = \mbox{``}p(s_{1},...,s_{n}) \from B\mbox{''} \in \P\}$}.

A substitution $\sigma$ is a function $\VV\rightarrow \TT(\Sigma,\VV)$ and we often denote its
application to a term $t$ by $t\sigma$ instead of $\sigma(t)$. As usual,
$Dom(\sigma) = \{ X \mid X\sigma \neq X\}$ and  $Range(\sigma) = \{
X\sigma \mid X \in Dom(\sigma) \}$. The restriction of $\sigma$ to  $\VV'
\subseteq \VV$  is  $\sigma\mbox{\scriptsize $|$}_{\VV'}(X) = \sigma(X)$ if $X \in
\VV'$, and $\sigma\mbox{\scriptsize $|$}_{\VV'}(X) = X$ otherwise. A substitution $\sigma$ is 
the \emph{most
general
unifier} (mgu) of  $s$ and $t$ iff $s\sigma = t\sigma$ and, whenever $s\gamma =
t\gamma$
for some $\gamma$, there exists a $\delta$ such that $X\gamma =
X\sigma\delta$ for all $X \in \VV(s) \cup \VV(t)$.
If $s$ and $t$
have no mgu, we write $s \not\sim t$.
Finally, to denote the term resulting from replacing all occurrences of a function symbol $f$ in a term $t$ by another
function symbol $g$, we write $t[f/g]$.

Now we recapitulate the operational semantics of logic programming with cut. Compared to
other formulations like
\cite{DBLP:journals/tplp/Andrews03,DBLP:journals/tcs/Billaud90,DBLP:journals/scp/Vink89,KulasBeierle,DBLP:journals/jlp/Spoto00},
the advantage of our
formalization is that it is particularly suitable for an extension to \emph{classes} of
queries in \rSC{sec:pet:abstract} and \ref{sec:pet:termgraph}, and  for synthesizing
cut-free programs in \rSC{sec:pet:graphs}. A formal
proof on the correspondence of our inference rules to the semantics of
the \textsf{Prolog} ISO standard \cite{ISOProlog} can be found in \cite{StroederDA}.

Our semantics is given by 7 inference rules.
They operate on \emph{states} which represent
the current goal, and also the
backtrack information that is needed to describe the effect of cuts.
The backtrack information is
given by a sequence of goals which are optionally labeled by the program clause that has to be applied to the
goal next. Moreover, our states also contain explicit \emph{marks} for the scope of a cut.

\begin{definition}[Concrete State]
A \emph{concrete state} is a sequence of elements from
\mbox{\small $\Goal(\Sigma,\VV) \cup (\Goal(\Sigma,\VV) \times \N \times \N) \cup {}$}\linebreak
\mbox{\small $\{ ?_n
\mid n \in \N\}$}, where  elements are separated by ``$\mid$''.
\mbox{\small $\State(\Sigma,\VV)$} is the set of all states.
\end{definition}

So an
element of a state
can be  $Q \in \Goal(\Sigma,\VV)$; or a labeled
goal $Q^{i}_{m} \in \Goal(\Sigma,\VV) \times \N \times \N$ representing that
we must apply the $i$-th program clause to  $Q$ next, where $m$ determines how a
cut introduced by the $i$-th clause will be labeled; or
$?_m$.
Here, $?_m$ serves as a marker to denote the end of the scope of cuts $!_m$
labeled with $m$.
Whenever a cut  $!_m$ is reached, all
elements preceding $?_m$ are discarded.

Now we express derivations in logic programming with cut
by seven  rules. Here,  $S$ and $S'$ are concrete states and
the goal $Q$ may also be $\Box$ (then ``$t,Q$'' is $t$).

\begin{definition}[Semantics with Concrete Inference Rules]\label{pet:def:concrete rules}

\vspace*{-.1cm}

{\footnotesize \begin{mathpar}
\inferrule*[right=(\textsc{Suc})\index{rule!Success@\textsc{Success}}]{\Box \mid S}{S}
\hspace*{-.2cm} \and\hspace*{-.2cm}
\inferrule*[right=(\textsc{Fail})\index{rule!Failure@\textsc{Fail}}]{?_m \mid S}{S}
\hspace*{-.2cm}\and\hspace*{-.2cm}
\inferrule*[right=(\textsc{Cut})\index{rule!Cut@\textsc{Cut}}]{!_m, Q \mid S \mid\,?_m \mid S'}{Q \mid\,?_m \mid S'}
\;\begin{minipage}{1cm}\vspace*{-3ex} where $S$ contains no $?_m$\end{minipage}
\hspace*{-.2cm}\and\hspace*{-.2cm}
\inferrule*[right=(\textsc{Cut})]{!_m, Q \mid
S}{Q}\;\begin{minipage}{1cm}\vspace*{-3ex}where $S$ contains no
$?_m$\end{minipage}
\end{mathpar}}

\vspace*{-.6cm}

\begin{mathpar}
\inferrule*[right=(\textsc{Case})\index{rule!Case@\textsc{Case}}]{t,Q \mid S}{(t,Q)^{i_1}_m \mid \ldots \mid (t,Q)^{i_k}_m \mid\,?_m \mid S}
\quad\begin{minipage}{6.1cm}\vspace*{-.5cm}where $t$ is neither a cut nor a variable, $m$ is greater than all previous marks,  and
$\Slice(\PP,t) = \{c_{i_1}, \ldots, c_{i_k}\}$ with $i_1 < \ldots < i_k$\end{minipage}
\end{mathpar}

\vspace*{-.4cm}

\begin{mathpar}
\inferrule*[right=(\textsc{Eval})\index{rule!Eval@\textsc{Eval}}]{(t,Q)^i_m
\mid S}{B_i'\sigma,Q\sigma \mid
S}\;\begin{minipage}{2.4cm}\vspace*{-3ex}where\\$c_i = H_i \from B_i$,\\$mgu(t,H_i)\!=\!\sigma$,\\$B_i' = B_i[!~/~!_m]$.
\end{minipage}
\hspace*{-.1cm}\and\hspace*{-.1cm}
\inferrule*[right=(\textsc{Backtrack})\index{rule!Backtrack@\textsc{Backtrack}}]{(t,Q)^i_m
\mid S}{S}\;\begin{minipage}{2.3cm}\vspace*{-3ex}where\\$c_i = H_i\from B_i$\\and $t~\not\sim~H_i$.
\end{minipage}
\end{mathpar}
\end{definition}

The \textsc{Suc} rule
is applicable if the first goal of our
sequence could be proved.
As we handle universal termination, we then have to backtrack to the next goal
in the sequence.
\textsc{Fail} means that for the current \mbox{$m$-th} case analysis, there
are no further backtracking
possibilities.
But the whole derivation does not have to fail,
 since the state
$S$ may still contain further alternative goals which have to be examined.

To make the backtracking possibilities explicit, the resolution of a
program
clause with the first atom $t$ of the current goal is split into two
operations.
 The \textsc{Case} analysis determines which clauses could
be applied to $t$ by slicing the
program according to $t$'s root symbol. It replaces the
current goal $(t,Q)$ by a goal labeled with the index $i_1$ of the first such clause and
adds copies of $(t,Q)$
labeled by the indices $i_2,\ldots,i_k$ of the other potentially applicable
clauses as backtracking possibilities. Note that
here, the top-down clause selection rule is taken into account.
Additionally, these goals are labeled
by a fresh mark $m \in \N$ that is greater than all previous marks, and
$?_m$ is added
at the end of the new backtracking goals
to denote
\begin{wrapfigure}[7]{r}{0.55\textwidth}
\vspace*{-.51cm}
{\tiny
  \hspace*{-.3cm} \begin{tikzpicture}
        [node/.style={rectangle,draw=blue!50,fill=blue!20,thick,inner sep=5pt},
        pre/.style={<-,thick}]
        \node[node] (div1) {$\Fdiv(\Fz,\Fz,Z)$};
        \node[node] (div2)  [right=.8 of div1] {$\Fdiv(\Fz,\Fz,Z)^{\tiny \ref{ex:divminus:1}}_1 \mid \Fdiv(\Fz,\Fz,Z)^{\tiny \ref{ex:divminus:2}}_1 \mid \Fdiv(\Fz,\Fz,Z)^{\tiny \ref{ex:divminus:3}}_1 \mid\,?_1$}
          edge [pre] node[auto,swap] {\textsc{Case}} (div1);
        \node[node] (div3)  [below right=.9 of div1, xshift=-2.12cm] {$!_1, \Ffailure(\Fa) \mid \Fdiv(\Fz,\Fz,Z)^{\tiny \ref{ex:divminus:2}}_1 \mid \Fdiv(\Fz,\Fz,Z)^{\tiny \ref{ex:divminus:3}}_1 \mid\,?_1$}
          edge [pre] node[auto,swap] {\textsc{Eval}}  node[auto] {$id$} (div2);
        \node[node] (div4)  [right=.8 of div3] {$\Ffailure(\Fa) \mid\,?_1$}
          edge [pre] node[auto,swap] {\textsc{Cut}} (div3);
  \node[node] (div4a)  [below right=.7 of div3, xshift=-5.1cm] {$\Ffailure(\Fa)^{\tiny \ref{ex:fail}}_2
\mid\,?_2 \mid\,?_1$}
          edge [pre] node[auto,xshift=-.36cm,yshift=-.15cm] {\textsc{Case}} (div4);
        \node[node] (div5)  [right=1.3 of div4a] {$?_2 \mid\,?_1$}
          edge [pre] node[auto,swap] {\textsc{Backtrack}} (div4a);
        \node[node] (div6)  [right=.6 of div5] {$?_1$}
          edge [pre] node[auto,swap] {\textsc{Fail}} (div5);
        \node[node] (div7)  [right=.6 of div6] {$\varepsilon$}
          edge [pre] node[auto,swap] {\textsc{Fail}} (div6);
  \end{tikzpicture}}
\end{wrapfigure}
the scope
of cuts.
For instance, consider the program  of \rEX{ex:divminus} and the
query $\Fdiv(\Fz,\Fz,Z)$. Here, we obtain the sequence depicted at
the side. The \textsc{Case} rule results in a state which  represents a case
analysis  where we first
try to apply the first $\Fdiv$-clause (\ref{ex:divminus:1}). When backtracking later on, we
use
clauses (\ref{ex:divminus:2}) and (\ref{ex:divminus:3}).

For a goal $(t,Q)^i_m$,
if  $t$
unifies with the head $H_i$ of the corresponding clause, we apply \textsc{Eval}. This rule
replaces $t$ by the body $B_i$ of the
clause and applies the mgu $\sigma$ to the result. When depicting rule applications as trees,
the corresponding
edge is labeled with
$\sigma|_{\VV(t)}$.
All cuts occurring in $B_i$ are labeled with $m$. The reason is
that if one reaches such a cut, then all further alternative goals up
to $?_m$ are discarded.

If $t$ does not unify with $H_i$, we apply the \textsc{Backtrack} rule.
Then, Clause $i$
cannot be used and we just backtrack to the next possibility in our backtracking
sequence.

Finally, there are two \textsc{Cut} rules. The first rule
removes all backtracking information on the level $m$ where the cut was
introduced. Since the explicit scope is represented by
$!_m$ and $?_m$, we have turned the cut into a \emph{local}
operation depending solely on the current state. Note
that $?_m$
must not be deleted as the
current goal $Q$ could still
lead to
another
cut $!_m$.
 The second \textsc{Cut} rule is used if $?_m$ is missing (e.g.,
if a cut $!_m$ is
already in the initial query). Later on, such states can also result from the
additional
\textsc{Parallel} inference rule which
will be introduced in \rSC{sec:pet:termgraph}. We treat such states as if $?_m$ were added
at the end of the backtracking sequence.

Note that these rules do not overlap, i.e., there is at most one rule that can be applied to any state.
The only case where no rule is applicable is when the state is the empty sequence
(denoted $\varepsilon$) or when the first goal starts with a variable.

The rules of \rDF{pet:def:concrete rules}
define the semantics of logic programs with cut using states. They
can also be
used to define the semantics using derivations between goals:
there is a derivation from the goal $Q$ to  $Q'$
in the program $\PP$ (denoted $Q \vdash^*_{\PP,\theta} Q'$) iff repeated application of our
rules can transform the state\footnote{If $Q$ contains cuts, then the inference rules
have to be applied to $Q[!/!_1]$ instead of $Q$.} $Q$ to a
state of the form
$\overline{Q}' \mid S$
for some  $S$, and $Q'$ results from $\overline{Q}'$ by removing all labels. Moreover, $\theta = \theta_1
\theta_2 \ldots \theta_n$ where $\theta_1,\ldots,\theta_n$ are the mgu's used in those
applications of the \textsc{Eval} rule that led to  $\overline{Q}'$. We call
$\theta\mbox{\scriptsize $|$}_{\VV(Q)}$ the corresponding \emph{answer substitution}. If $\theta$ is not of
interest, we write $\vdash_\PP$ instead of $\vdash_{\PP,\theta}$.

Consequently, our inference rules can be used for termination proofs: If there is an infinite
derivation (w.r.t.\ $\vdash_\PP$) starting in some goal $Q$, then there is also an infinite
sequence of inference rule applications starting in the state $Q$,
i.e., $Q$ is a ``non-terminating state''. Note that we distinguish derivations in logic
programming (i.e., $Q \vdash_\PP Q'$ for goals $Q$ and $Q'$) from sequences of states that result
from application of the inference rules in \rDF{pet:def:concrete rules}. If a state $S$ can be
transformed 
into a state $S'$ by such an inference rule, we speak of a
``\emph{state-derivation}''.

\section{Abstract Derivations}\label{sec:pet:abstract}

To represent \emph{classes} of queries, we introduce \emph{abstract terms}
 and a set  $\AA$ of
\emph{abstract variables}, where each $T \in \AA$ represents a fixed but
arbitrary term. $\NN$ consists of all ``ordinary''
variables  in logic programming. Then, as \emph{abstract terms} we
consider all terms from the set $\TT(\Sigma,\VV)$ where $\VV = \NN \uplus
\AA$. \emph{Concrete terms} are terms from $\TT(\Sigma,\NN)$, i.e., terms
containing no abstract variables. For any set  $\VV' \subseteq
\VV$, let $\VV'(t)$ be the variables from $\VV'$ occurring in the term $t$.

To 
determine by which terms an abstract variable may be
instantiated,
we add
a knowledge base $\mathit{KB} = (\GG,
\UU)$ to each state,
where $\GG \subseteq \AA$ and
 $\UU \subseteq \TT(\Sigma,\VV) \times
\TT(\Sigma,\VV)$.
The variables in $\GG$ may only be instantiated by
ground terms.
And  $(s,s') \in \UU$ means that we are restricted to
instantiations $\gamma$ of the abstract variables where
$s\gamma \not\sim s'\gamma$, i.e., $s$ and $s'$ may not become unifiable when
instantiating them with $\gamma$.

\begin{definition}[Abstract State]
The set of \emph{abstract states}\index{state!abstract} $\AState(\Sigma,\NN,\AA)$ is a set of pairs $(S; \mathit{KB})$
of a concrete state $S \in \State(\Sigma,\NN \cup \AA)$ and a \emph{knowledge
base}\index{state!knowledge base} $\mathit{KB}$.
\end{definition}

A substitution $\gamma$ is  a
\emph{concretization} of an abstract state if it respects
the knowledge base $(\GG,
\UU)$. So first, $\gamma$
instantiates all abstract variables, i.e.,
$Dom(\gamma) = \AA$. Second, when applying $\gamma$,
the resulting term must be concrete, i.e.,
$\VV(Range(\gamma)) \subseteq \NN$. Third,
abstract variables from $\GG$ may
only be replaced by ground terms, i.e.,
$\VV(Range(\gamma\mbox{\scriptsize $|$}_{\GG})) = \emptyset$.
Fourth, for all pairs $(s,s') \in \UU$, $s\gamma$ and $s'\gamma$ must not unify.

\begin{definition}[Concretization]
A substitution $\gamma$ is a
\emph{concretization}\index{concretization} w.r.t.\  $(\GG,
\UU)$
iff $Dom(\gamma) = \AA$, $\VV(Range(\gamma)) \subseteq \NN$,
$\VV(Range(\gamma\mbox{\scriptsize $|$}_{\GG})) = \emptyset$, and
$s\gamma \not\sim s'\gamma$ for all $(s,s') \in \UU$.
The set \pagebreak of concretizations
of an abstract state $(S; \mathit{KB})$ is $\SSS(S; \mathit{KB}) =
\{ S\gamma \mid \gamma\text{ is a concretization w.r.t. } \mathit{KB}\}$.
\end{definition}

\begin{example}
\label{pet:ex:concretes}
Consider the abstract state which consists of the single goal $\Fminus(T_1,
T_2, T_3)$ and the knowledge base $(\{T_1, T_2\},
\{(T_1,T_3)\})$, with
 $T_{i} \in \AA$ for all $i$. So here $\GG = \{T_1, T_2\}$ and $\UU$ only contains
$(T_1,T_3)$.
This represents all concrete states $\Fminus(t_1, t_2, t_3)$ where $t_1, t_2$ are ground
terms and where $t_1$ and $t_3$ do not unify, i.e., $t_3$ does not match $t_1$.
For example,  $\Fminus(\Fz, \Fz, Z)$ is not represented as $\Fz$ and
$Z$ unify. In contrast,  $\Fminus(\Fs(\Fz),\Fs(\Fz),\Fz)$ and $\Fminus(\Fz,\Fz,\Fs(\Fz))$
are represented. Note that $\Fminus(\Fs(\Fz),\Fs(\Fz),\Fz)$
 can be reduced to
$\Fminus(\Fz,\Fz,\Fz)$ using Clause
(\ref{ex:divminus:6}) from \rEX{ex:divminus}. But  Clause
(\ref{ex:divminus:6}) cannot be
applied to all concretizations. For example, the concrete state $\Fminus(\Fz,\Fz,\Fs(\Fz))$
is also represented by our abstract state, but here  no clause is applicable.
\end{example}

\rEX{pet:ex:concretes} demonstrates that we need to adapt our inference rules to
reflect that sometimes a clause can be applied only for some concretizations
of the abstract variables, and to exploit the
information from the knowledge base of the abstract state.
We now adapt our inference rules to abstract states that represent \emph{sets} of concrete
states. The
invariant of our rules is that all states represented by the parent node are terminating
if all the states represented by its children are terminating.

\begin{definition}[Sound Rules]
An abstract state is called \emph{terminating} iff all its
concretizations are terminating. A rule $\rho: \AState(\Sigma,\NN,\AA) \to
2^{\AState(\Sigma,\NN,\AA)}$ is \emph{sound}
if  $(S; \mathit{KB})$ is terminating whenever
all $(S'; \mathit{KB}') \in \rho(S; \mathit{KB})$ are
terminating.
\end{definition}

The rules \textsc{Suc}, \textsc{Fail}, \textsc{Cut}, and \textsc{Case} do not change
the knowledge base and are, thus, straightforward to adapt.
Here,
 $S \mid S'; \mathit{KB}$ stands for  $((S \mid S'); \mathit{KB})$.

\begin{definition}[Abstract Inference Rules -- Part 1 (\textsc{Suc},
\textsc{Fail}, \textsc{Cut}, \textsc{Case})] 
\label{pet:def:abstract rules 1}

\vspace*{-.4cm}

\begin{mathpar}
\inferrule*[right=(\textsc{Suc})\index{rule!Success@\textsc{Suc}}]{\Box \mid S; \mathit{KB}}{S; \mathit{KB}}
\and
\inferrule*[right=(\textsc{Fail})\index{rule!Failure@\textsc{Fail}}]{?_m \mid
S; \mathit{KB}}{S; \mathit{KB}} 
\\
\and
\inferrule*[right=(\textsc{Cut})\index{rule!Cut@\textsc{Cut}}]{!_m, Q \mid S \mid\,?_m \mid S'; \mathit{KB}}{Q \mid\,?_m \mid S'; \mathit{KB}}\ \begin{minipage}{1.5cm}\vspace*{-3ex}where $S$ contains no $?_m$\end{minipage}
\and
\inferrule*[right=(\textsc{Cut})]{!_m, Q \mid S; \mathit{KB}}{Q; \mathit{KB}}\ \begin{minipage}{1.5cm}\vspace*{-3ex}where $S$ contains no $?_m$\end{minipage}
\and
\inferrule*[right=(\textsc{Case})\index{rule!Case@\textsc{Case}}]{t,Q \mid S; \mathit{KB}}{(t,Q)^{i_1}_m \mid \ldots \mid (t,Q)^{i_k}_m \mid\,?_m \mid S; \mathit{KB}}
\quad\begin{minipage}[t]{5.5cm}\vspace*{-.9cm}where $t$ is neither a cut nor a variable,
$m$ is greater than all previous marks,  and $\Slice(\PP,t) = \{c_{i_1}, \ldots, c_{i_k}\}$ with $i_1 < \ldots <
i_k$\end{minipage}
\end{mathpar}
\end{definition}

In \rDF{pet:def:concrete rules},
we determined which of the rules \textsc{Eval} and
\textsc{Backtrack}
to apply by trying to unify the first
atom $t$ with the head $H_i$ of the corresponding clause. But as demonstrated by
\rEX{pet:ex:concretes}, in the abstract case we might need to apply
\textsc{Eval} for some concretizations
and \textsc{Backtrack} for others. \textsc{Backtrack} can be used for \emph{all}
concretizations
if $t$ does not
unify with $H_{i}$ or if their mgu
contradicts $\UU$.
 This gives rise to the abstract \textsc{Backtrack} rule in the following
definition.
When the abstract \textsc{Backtrack} rule is not applicable, we still cannot be sure that
$t\gamma$ unifies with $H_i$ for all concretizations $\gamma$.
Thus,  we have an abstract \textsc{Eval}
rule  with two successor states that combines both
the concrete \textsc{Eval} and the concrete \textsc{Backtrack} rule.

\begin{definition}[Abstract Inference Rules -- Part 2 (\textsc{Backtrack}, \textsc{Eval})]
\label{pet:def:abstract rules 3}

\vspace*{-.3cm}

\begin{mathpar}
\inferrule*[right=(\textsc{Backtrack})\index{rule!Backtrack@\textsc{Backtrack}}]{(t,Q)^i_m
\mid S; \mathit{KB}}{S; \mathit{KB}}
\quad
\begin{minipage}[t]{7cm}
\vspace*{-.7cm}
where $c_i = H_i \from B_i$ and
there is no concretization $\gamma$
w.r.t.\ $\mathit{KB}$
such that  $t\gamma \sim H_i$.
\end{minipage}
\end{mathpar}
\vspace*{-.5cm}
\begin{mathpar}
\inferrule*[right=(\textsc{Eval})\index{rule!Eval@\textsc{Eval}}]{(t,Q)^i_m \mid S; (\GG, 
\UU)}{B_i'\sigma,Q\sigma \mid S\sigma\mbox{\scriptsize $|$}_\GG; (\GG', 
\UU\sigma\mbox{\scriptsize $|$}_\GG)\\S; (\GG,
\UU \cup \{(t,H_i)\})}
\end{mathpar}
where $c_i = H_i \from B_i$ and $mgu(t,H_i) = \sigma$. W.l.o.g.,
$\VV(\sigma(X))$ only contains fresh abstract variables for all $X \in \VV$.
Moreover,   $\GG' =
\AA(Range(\sigma\mbox{\scriptsize $|$}_\GG))$ and $B_i' = B_i[! / !_m]$.
\end{definition}

In \textsc{Eval}, w.l.o.g.\
we assume that  $mgu(t,H_i)$ renames all variables to fresh abstract
variables. This is needed to handle ``sharing'' effects correctly, i.e., to handle
concretizations  which introduce multiple occurrences of (concrete) variables, cf.\ \cite{techreport}.
The knowledge base is updated differently
for the successors corresponding to the concrete
\textsc{Eval}  and  \textsc{Backtrack} rule.
For all
concretizations corresponding to the second successor of \textsc{Eval},
the concretization of $t$ does not unify with $H_i$. Hence, here we
add the pair $(t, H_i)$ to the set $\UU$.

Now consider
concretizations $\gamma$ where $t\gamma$ and
$H_i$ unify, i.e., concretizations $\gamma$ corresponding to the first successor of
the \textsc{Eval} rule. Then for any  $T \in \GG$, $T\gamma$
is a ground instance of $T\sigma$. Hence, we  replace
all  $T \in \GG$ by  $T\sigma$, i.e., we apply $\sigma\mbox{\scriptsize $|$}_\GG$ to $\UU$ and
$S$.
Now the new set
$\GG'$ of abstract variables that may only be
instantiated by ground terms is $\AA(Range(\sigma\mbox{\scriptsize $|$}_{\GG}))$.
As before, $t$ is replaced by the instantiated clause body $B_i$ where we label cuts
 with the
number $m$ of the current \textsc{Case} analysis.

Now any concrete derivation with the rules from \rDF{pet:def:concrete
rules} can also be simulated with the abstract rules from \rDF{pet:def:abstract rules 1}
and \ref{pet:def:abstract rules 3}. But unfortunately, even for terminating goals,
in general these rules yield
an infinite tree.   The reason is that  there is no bound on  the size of terms
 represented by
the abstract variables and hence,
\begin{wrapfigure}[14]{r}{0.6\textwidth}
\vspace*{-.4cm}
{\tiny
\begin{tikzpicture}
[node/.style={rectangle,draw=blue!50,fill=blue!20,thick,inner sep=5pt},
pre/.style={<-,thick}]
\begin{scope}[node distance=0.4 and 0.4]
\node[node] (div1) {$\Fp(T_1); (\{T_1\},
\emptyset)$};
\node[node] (div2)  [below=of div1] {$\Fp(T_1)^{\tiny \ref{ex:p:1}}_1 \mid \,?_{1}; (\{T_1\}, 
\emptyset)$}
  edge [pre] node[auto,swap] {\textsc{Case}} (div1);
\node[node] (div3')  [below right=of div2] {$?_{1}; (\{T_1\}, 
\{(\Fp(T_1),\Fp(\Fs(X)))\})$}
  edge [pre] node[auto,swap] {\textsc{Eval}} (div2);
\node[node] (div4')  [below=of div3'] {$\varepsilon; (\{T_1\}, 
\{(\Fp(T_1),\Fp(\Fs(X)))\})$}
  edge [pre] node[auto,swap] {\textsc{Fail}} (div3');
\node[node] (div3)  [below=of div2] {$\Fp(T_2) \mid\,?_{1}; (\{T_2\}, 
\emptyset)$}
  edge [pre] node[auto,swap] {\textsc{Eval}} node[auto] {$T_1/\Fs(T_2)$} (div2);
\node[node] (div4)  [below=of div3] {$\Fp(T_2)^{\tiny \ref{ex:p:1}}_2 \mid\,?_{2} \mid\,?_{1}; (\{T_2\},
\emptyset)$}
  edge [pre] node[auto,swap] {\textsc{Case}} (div3);
\node[node] (div5')  [below right=of div4] {$?_{2} \mid\,?_{1}; (\{T_2\}, 
\{(\Fp(T_2),\Fp(\Fs(X)))\})$}
  edge [pre] node[auto,swap] {\textsc{Eval}} (div4);
\node[node] (div5)  [below=of div4] {$\Fp(T_3) \mid\,?_{2} \mid\,?_{1}; (\{T_3\}, 
\emptyset)$}
  edge [pre] node[auto,swap] {\textsc{Eval}} node[auto] {$T_2/\Fs(T_3)$} (div4);
\node[inner sep=5pt] (div6)  [below=of div5] {$\ldots$}
  edge [pre] node[auto,swap] {\textsc{Case}} (div5);
\node[inner sep=5pt] (div6')  [below=of div5'] {$\ldots$}
  edge [pre] node[auto,swap] {\textsc{Fail}} (div5');
\end{scope}
\end{tikzpicture}
}
\end{wrapfigure}
the abstract \textsc{Eval} rule can be
applied infinitely often.

\begin{example}
\label{ex:p}
Consider the 1-rule
program

\vspace*{-.3cm}

{\small \begin{equation}
\label{ex:p:1}\Fp(\Fs(X))  \from  \Fp(X).
\end{equation}}

\vspace*{-.4cm}

\noindent
For queries of the form $\Fp(t)$ where $t$ is ground, the program terminates.
However, the tree built using the abstract inference rules is obviously infinite.
\end{example}

\section{From Trees to Graphs}\label{sec:pet:termgraph}

To obtain a finite graph instead of an infinite tree, we now
introduce
an additional \textsc{Instance} rule which
 allows us to connect the current state $(S; \mathit{KB})$
with a previous state $(S'; \mathit{KB}')$, provided that the current state is an instance of the previous
state. In other words,  every concretization of  $(S;\mathit{KB})$ must 
be a
concretization of  $(S'; \mathit{KB}')$.
Still, \textsc{Instance} is often
not enough to obtain a \pagebreak finite graph.

\addtocounter{example}{1}
\setcounter{ex:pp:version1}{\value{example}}

\vspace*{.2cm}

\noindent
\textit{Example \arabic{ex:pp:version1}}

\noindent
We extend \rEX{ex:p} by the following additional
fact.
\begin{equation}
\label{ex:p:2:ref}
  \Fp(X).
\end{equation}
For queries $\Fp(t)$
 where $t$ is ground, the  program still
terminates. If we start with
$(\Fp(T_1);
(\{T_1\},\varnothing))$, then the \textsc{Case} rule results in the
state $(\Fp(T_1)^{\footnotesize \ref{ex:p:1}}_1 \mid \Fp(T_1)^{\footnotesize \ref{ex:p:2:ref}}_1 \mid\; ?_1;\linebreak
(\{T_1\},\varnothing))$ and the \textsc{Eval} rule produces two new states,
one of them being  $(\Fp(T_2) \mid \Fp(\Fs(T_2))^{\footnotesize \ref{ex:p:2:ref}}_1 \mid\; ?_1;
(\{T_2\},\varnothing))$.

To simplify states,
from now on we will eliminate so-called \emph{non-active}
marks $?_m$ which occur as first or as last element
in states. Eliminating $?_m$ from the beginning of a state is possible,
as  \textsc{Fail} would also remove such a
$?_{m}$. Eliminating $?_m$ from the end of a
state is possible, as applying the first \textsc{Cut}
rule to a state ending in $?_{m}$ is equivalent
to applying the second \textsc{Cut} rule to the same state without $?_{m}$.

We will
also reduce the knowledge
base to just those abstract variables that occur in
the state and remove pairs $(s,s')$
from $\UU$ where $s  \not\sim s'$.
Still, $(\Fp(T_2) \mid$
\begin{wrapfigure}[9]{r}{0.58\textwidth}
\vspace*{-.3cm}
{\tiny
\hspace*{-.4cm}\begin{tikzpicture}
[node/.style={rectangle,draw=blue!50,fill=blue!20,thick,inner sep=5pt},
pre/.style={<-,thick},post/.style={->,thick,dashed}]
\begin{scope}[node distance=0.5 and 0.5]
\node[node] (div1) {$\Fp(T_1); (\{T_1\}, 
\emptyset)$};
\node[node] (div2)  [below=of div1] {$\Fp(T_1)^{\tiny \ref{ex:p:1}}_1 \mid \Fp(T_1)^{\ref{ex:p:2:ref}}_1; (\{T_1\}, 
\emptyset)$}
  edge [pre] node[auto,swap] {\textsc{Case}} (div1);
\node[node] (div3')  [right=1.3 of div2] {$\Fp(T_1)^{\ref{ex:p:2:ref}}_1; (\{T_1\}, 
\emptyset)$}
  edge [pre] node[auto,swap] {\textsc{Parallel}} (div2);
\node[node] (div3)  [below=of div2] {$\Fp(T_1)^{\tiny \ref{ex:p:1}}_1; (\{T_1\}, 
\emptyset)$}
  edge [pre] node[auto,swap] {\textsc{Parallel}} (div2);
\node[node] (div4)  [below=of div3] {$\Fp(T_2); (\{T_2\},
\emptyset)$}
  edge [pre] node[auto,swap] {\textsc{Eval}} node[auto] {$T_1/\Fs(T_2)$} (div3)
  edge [post,out=180,in=180] node[auto,swap] {\textsc{Instance}} (div1);
\node[node] (div5)  [below right=of div3,xshift=-.5cm] {$\varepsilon; (\emptyset,
\emptyset)$}
  edge [pre] node[auto,swap] {\textsc{Eval}} (div3);
\node[node] (div4')  [below left=of div3',xshift=.8cm] {$\Box; (\emptyset,
\emptyset)$}
  edge [pre] node[auto,swap,yshift=.1cm] {\textsc{Eval}} node[auto,yshift=-.1cm] {$T_1 / T_2$} (div3');
\node[node] (div4'')  [below=of div3'] {$\varepsilon; (\emptyset,
\emptyset)$}
  edge [pre] node[auto,swap] {\textsc{Eval}} (div3');
\node[node] (div4'')  [below right=of div4',xshift=-1cm,yshift=-.08cm] {$\varepsilon; (\emptyset,
\emptyset)$}
  edge [pre] node[auto,swap] {\textsc{Suc}} (div4');
\end{scope}
\end{tikzpicture}}
\end{wrapfigure}
$\Fp(\Fs(T_2))^{\footnotesize \ref{ex:p:2:ref}}_1;
(\{T_2\},\varnothing))$
is not an instance of the previous state
$(\Fp(T_1);(\{T_1\},\varnothing))$ due to the ad-\linebreak ded backtrack goal
$\Fp(\Fs(T_2))^{\footnotesize \ref{ex:p:2:ref}}_1$.
Therefore, we
now introduce a
\textsc{Parallel} rule that allows us to split a backtracking sequence into
separate problems.
Now
we obtain the graph on the right.

\vspace*{.2cm}

Clearly,
 \textsc{Parallel} may
transform terminating into non-termina\-ting states.
But without further conditions, \textsc{Parallel}
is not only ``incomplete'', but also
unsound. Consider a state
$(\,!_2 \mid\,
!_1 \mid\, ?_2 \mid \Fp; (\varnothing,\varnothing))$ for the program  $\Fp
\from \Fp$. The state is not terminating, as $!_1$ is not
reachable. Thus, one eventually evaluates
$\Fp$. But if one splits the state into
$(!_2;  (\varnothing,\varnothing))$ and $(!_1 \mid ?_2 \mid \Fp;
(\varnothing,\varnothing))$, both new states terminate.

To solve this problem, in addition to the ``active marks'' (cf.\ Ex.\ \arabic{ex:pp:version1})
we introduce the notion of \emph{active cuts}. The active cuts of a state $S$ are
those $m \in \N$ where $!_{m}$ occurs in $S$ or where $!_{m}$ can be introduced
by \textsc{Eval} applied to a labeled goal $(t,q)^{i}_{m}$ occurring in
$S$.
Now the  \textsc{Parallel} rule may only split a backtracking sequence into two parts $S$ and
$S'$ if the active cuts of $S$ and the active marks of $S'$ are
disjoint.

\begin{definition}[Abstract Inference Rules -- Part 3 (\textsc{Instance}, \textsc{Parallel})]
\label{pet:def:abstract rules 4}

\vspace*{-.4cm}

\begin{mathpar}
\inferrule{S; (\GG,
\UU)}{S'; (\GG',
\UU')}\
(\textsc{Instance})\index{rule!Instance@\textsc{Instance}}\quad\begin{minipage}{8.2cm}if
there is a $\mu$ such that $S = S'\mu$,  $\mu|_\NN$ is a variable renaming,
$\VV(T\mu) \subseteq \GG$
for all $T \in \GG'$,  
and $\UU'\mu \subseteq \UU$.
\end{minipage}
\and
\inferrule*[right=(\textsc{Parallel})\index{rule!Parallel@\textsc{Parallel}}]{S \mid S'; \mathit{KB}}{S; \mathit{KB} \\ S'; \mathit{KB}}\quad\begin{minipage}{5cm}\vspace*{-3ex}if $AC(S) \cap AM(S') = \emptyset$\end{minipage}
\end{mathpar}
The \emph{active cuts} $AC(S)$ are
all $m$ where $!_m$ is in $S$ or $(t,q)^i_m$ is in $S$ and $c_i$'s body has a cut.
The \emph{active marks} $AM(S)$ are all $m$ where
{\small $S = S'\mid\,?_m \mid S''$} and {\small $S' \neq \varepsilon, S'' \neq
\varepsilon$}.
\end{definition}

\addtocounter{example}{1}
\setcounter{ex:pq}{\value{example}}

\begin{wrapfigure}[11]{r}{0.37\textwidth}
\vspace*{-.1cm}
{\tiny
\begin{tikzpicture}
[node/.style={rectangle,draw=blue!50,fill=blue!20,thick,inner sep=5pt},
pre/.style={<-,thick}]
\begin{scope}[node distance=0.45 and 0.45]
\node[node] (div1) {$\Fp(T_1); (\{T_1\},
\emptyset)$};
\node[node] (div2)  [below=of div1] {$\Fp(T_1)^{\tiny \ref{ex:pq:1}}_1; (\{T_1\}, 
\emptyset)$}
  edge [pre] node[auto,swap] {\textsc{Case}} (div1);
\node[node] (div3)  [below=of div2] {$\Fp(T_2), \Fq; (\{T_2\}, ,
\emptyset)$}
  edge [pre] node[auto,swap] {\textsc{Eval}} node[auto] {$T_1/\Fs(T_2)$} (div2);
\node[node] (div3')  [right=.7 of div3] {$\varepsilon; (\emptyset, 
\emptyset)$}
  edge [pre] node[auto,swap] {\textsc{Eval}} (div2);
\node[node] (div4)  [below=of div3] {$\Fp(T_2)^{\tiny \ref{ex:pq:1}}_2, \Fq; (\{T_2\}, 
\emptyset)$}
  edge [pre] node[auto,swap] {\textsc{Case}} (div3);
\node[node] (div5')  [below right=of div4] {$\varepsilon; (\emptyset, 
\emptyset)$}
  edge [pre] node[auto,swap] {\textsc{Eval}} (div4);
\node[node] (div5)  [below=of div4] {$\Fp(T_3), \Fq, \Fq; (\{T_3\}, 
\emptyset)$}
  edge [pre] node[auto,swap] {\textsc{Eval}} node[auto] {$T_2/\Fs(T_3)$} (div4);
\node[inner sep=5pt] (div6)  [below=of div5] {$\ldots$}
  edge [pre] node[auto,swap] {\textsc{Case}} (div5);
\end{scope}
\end{tikzpicture}}
\end{wrapfigure}
\noindent
\textit{Example \arabic{ex:pq}}

\noindent
However,
there are still examples where the graph cannot be ``closed''.
Consider the  program

\vspace*{-.2cm}

\hspace{-.4cm}
\parbox{120pt}{
\begin{equation}
\label{ex:pq:1}\Fp(\Fs(X))  \from  \Fp(X), \Fq.
\end{equation}
}
\hspace*{.8cm}
\parbox{40pt}{
\begin{equation}
\label{ex:pq:2}\Fq.
\end{equation}
}

\noindent
For queries $\Fp(t)$ where $t$ is ground, the  program again  terminates.
With  \rDF{pet:def:abstract rules 1},
\ref{pet:def:abstract rules 3}, and  \ref{pet:def:abstract rules 4},
we obtain the infinite tree on the right.
It never encounters an instance of a previous state, since
each resolution with  Clause (\ref{ex:pq:1})
adds a $\Fq$ to the goal.

\vspace*{.2cm}

Thus, we introduce a final abstract \textsc{Split}
rule to split a state $(t,Q; \mathit{KB})$
into  $(t; \mathit{KB})$  and a state $(Q\mu; \mathit{KB}')$, where $\mu$
approximates
the answer substitutions for $t$.
The edge from $(t,Q; \mathit{KB})$ to $(Q\mu; \mathit{KB}')$ is labeled with $\mu|_{\VV(t) \cup \VV(Q)}$.
To simplify the \textsc{Split} rule,
we only define it for backtracking sequences
of one element. To obtain such a sequence, we can use the \textsc{Parallel} rule.

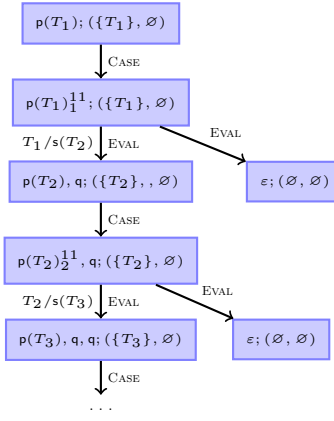
\begin{definition}[Abstract Inference Rules -- Part 4 (\textsc{Split})]
\label{pet:def:abstract rules 6}

\vspace*{-.3cm}

\begin{mathpar}
\inferrule*[right=(\textsc{Split})\index{rule!Split@\textsc{Split}}]{t,Q; (\GG,
\UU)}{ t; (\GG,
\UU) \\ Q\mu; (\GG',
\UU\mu)}
\quad\begin{minipage}{6cm}
\vspace*{-.7cm}
where $\mu$ replaces all variables from $\VV \setminus \GG$ by fresh abstract variables
and $\GG' = \GG \cup ApproxGnd(t,\mu)$.
\end{minipage}
\end{mathpar}

\noindent
Here, $ApproxGnd$\index{ApproxGnd@\textit{ApproxGnd}} is defined as follows.
We assume that we have a \emph{groundness analysis} function $Ground_\P : \Sigma
\times 2^\N \to 2^\N$, see, e.g., \cite{DBLP:journals/tplp/HoweK03}.
If $p$ is an $n$-ary predicate,
 $\{i_1,\ldots,i_m\} \subseteq \{1,\ldots,n\}$, and
$Ground_\P(p,\{i_1,\ldots,i_m\}) = \{j_1,\ldots,j_k\}$, then any successful derivation
$p(t_1,\ldots,t_n) \vdash^*_{\PP,\theta} \Box$
where
$t_{i_1},\ldots,t_{i_m}$ are ground will lead to an answer substitution
$\theta$ such that $t_{j_1}\theta, \ldots, t_{j_k}\theta$ are
ground. So $Ground_\P$ approximates which positions of $p$ will become ground
if the ``input'' positions  $i_1,\ldots,i_m$ are ground.
Now if $t= p(t_1,\ldots,t_n)$ is an abstract term where
$t_{i_1},\ldots,t_{i_m}$ are ground
in every concretization (i.e., all their variables are from $\GG$),
then
$ApproxGnd(t,\mu)$ returns the $\mu$-renamings of
all abstract variables that will be ground in every successful derivation starting from a
concretization of
$t$. Thus, $ApproxGnd(t,\mu)$
contains the abstract variables of $t_{j_1}\mu, \ldots, t_{j_k}\mu$. So formally
\[
ApproxGnd(p(t_1,\ldots,t_n), \mu) = \{\AA(t_j\mu) \mid  j \in Ground_{\PP}(p, \{i \mid \VV(t_i) \subseteq \GG\})\}
\]
\end{definition}

\addtocounter{example}{1}
\setcounter{div ill}{\value{example}}

\noindent
\textit{Example \arabic{div ill}}

\noindent
To illustrate \rDF{pet:def:abstract rules 6},
regard the
program of \rEX{ex:divminus} and the state  $(\Fminus(T_5, T_6, T_8),\linebreak
 \Fdiv(T_8, T_6, T_7);
(\{T_5, T_6 \}, \UU))$ with $T_5,T_6,T_7,T_8 \in \AA$.
(This state will occur in the
 termination proof of $\Fdiv$, cf.\
\rEX{ex:divminus:graphandlp}.)
 We have
$\GG = \{T_5, T_6 \}$ and hence if $\Fminus(t_1,t_2,t_3)$ is $\Fminus(T_5,
T_6, T_8)$,
then $Ground_{\PP}(\Fminus, \{i \mid \VV(t_i) \subseteq \GG\}) =
Ground_{\PP}(\Fminus, \{1,2\}) = \{1,2,3\}$.
In other words, if the first two arguments of $\Fminus$ are ground
and the derivation is successful, then the answer substitution
also instantiates the third argument to a ground term. Since $\mu$ only renames variables outside of $\GG$, we have
$\mu =\{ T_7/T_9, T_8/T_{10} \}$.  So  $ApproxGnd(\Fminus(T_5,T_6, T_8), \mu) = \{ \AA(t_1\mu), \AA(t_2\mu), \AA(t_3\mu) \}
=
\{ T_5\mu,
T_6\mu, T_8\mu \} = \{T_5, T_6, T_{10} \}$. So the \textsc{Split} rule transforms
the current state
\begin{wrapfigure}[11]{r}{0.52\textwidth}
\hspace*{-.3cm}{\tiny
\begin{tikzpicture}
[node/.style={rectangle,draw=blue!50,fill=blue!20,thick,inner sep=5pt},
pre/.style={<-,thick},post/.style={->,thick,dashed}]
\begin{scope}[node distance=0.5 and 0.5]
\node[node] (div1) {$\Fp(T_1); (\{T_1\}, 
\emptyset)$};
\node[node] (div2)  [below=of div1] {$\Fp(T_1)^{\tiny \ref{ex:pq:1}}_1; (\{T_1\}, 
\emptyset)$}
  edge [pre] node[auto,swap] {\textsc{Case}} (div1);
\node[node] (div3')  [right=of div2] {$\varepsilon; (\emptyset, 
\emptyset)$}
  edge [pre] node[auto,swap] {\textsc{Eval}} (div2);
\node[node] (div3)  [below=of div2] {$\Fp(T_2), \Fq; (\{T_2\}, 
\emptyset)$}
  edge [pre] node[auto,swap] {\textsc{Eval}} node[auto] {$T_1/\Fs(T_2)$} (div2);
\node[node] (div4)  [below=of div3] {$\Fp(T_2); (\{T_2\}, 
\emptyset)$}
  edge [pre] node[auto,swap] {\textsc{Split}} (div3)
  edge [post,out=180,in=180] node[auto,swap] {\textsc{Instance}} (div1);
\node[node] (div4')  [right=of div3] {$\Fq; (\emptyset, 
\emptyset)$}
  edge [pre] node[auto,swap] {\textsc{Split}} node[auto] {$id$} (div3);
\node[node] (div4'')  [below=of div4'] {$\Fq^{\tiny \ref{ex:pq:2}}_{2}; (\emptyset, 
\emptyset)$}
  edge [pre] node[auto,swap] {\textsc{Case}} (div4');
\node[node] (div5'')  [below right=of div4'',xshift=-.5cm] {$\varepsilon; (\emptyset, 
\emptyset)$}
  edge [pre] node[auto,swap] {\textsc{Eval}} (div4'');
\node[node] (div5')  [below=of div4''] {$\Box; (\emptyset, 
\emptyset)$}
  edge [pre] node[auto,swap] {\textsc{Eval}} node[auto] {$id$} (div4'');
\node[node] (div6')  [left=of div5'] {$\varepsilon; (\emptyset, 
\emptyset)$}
  edge [pre] node[auto,swap] {\textsc{Suc}} (div5');
\end{scope}
\end{tikzpicture}}
\end{wrapfigure}
to $(\Fminus(T_5, T_6, T_8); (\{T_5, T_6
\}, \UU))$
and  $(\Fdiv(T_{10}, T_6, T_9); \;
(\{T_5, T_6, T_{10} \}, \; \UU\mu))$\linebreak
 where one can eliminate $T_5$ from the new
groundness set $\GG'$.

With the additional \textsc{Split} rule,  we can always obtain finite
graphs instead of infinite trees. (This will be proved in \rTH{Existence of
Termination Graphs}.) Thus, no further rules are needed.
As depicted on the right,
now we can also close the graph for
Ex.\ \arabic{ex:pq}'s program.

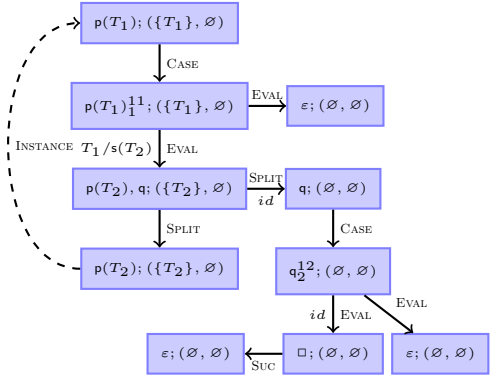
\vspace*{.1cm}

\rTH{soundness thm} proves the soundness of all our abstract inference
rules. In other
words, if all children of a node are terminating, then the node is terminating
as well.

\begin{theorem}[Soundness of the Abstract Inference Rules]\label{soundness thm}
The inference rules from \rDF{pet:def:abstract rules 1},
\ref{pet:def:abstract rules 3}, \ref{pet:def:abstract rules 4}, and
\ref{pet:def:abstract rules 6} are sound.\footnote{For all proofs, we refer to
\cite{techreport}.}
\end{theorem}

\section{From Termination Graphs to Logic Programs}\label{sec:pet:graphs}

Now we  introduce \emph{termination graphs} as a  subclass of
the graphs obtained by   \rDF{pet:def:abstract rules 1},
\ref{pet:def:abstract rules 3},  \ref{pet:def:abstract rules 4},
\ref{pet:def:abstract rules 6}. Then we show
how to extract cut-free  programs from termination graphs.

\begin{definition}[Termination Graph]
A finite graph built from an initial state $(S;\mathit{KB})$ using   \rDF{pet:def:abstract rules 1},
\ref{pet:def:abstract rules 3},  \ref{pet:def:abstract rules 4}, and
\ref{pet:def:abstract rules 6}
is a \emph{termination graph} iff there is no cycle consisting only of
\textsc{Instance} edges
 and
all leaves are of the form $(\varepsilon; \mathit{KB}')$ or
$(X,Q\mid S; \mathit{KB}')$ with $X \in \VV$.
If there are no leaves of the form $(X,Q\mid S; \mathit{KB}')$, then the graph is ``\emph{proper}''.
\end{definition}

We want to generate
clauses for the loops in the termination graph and show their termination.
Thus, there should be no cycles consisting only of \textsc{Instance} edges,
as\linebreak they would lead to trivially
non-terminating clauses.
Moreover, the only leaves  may be nodes where no inference rule is
applicable anymore (i.e., the graph must be ``fully expanded'').
For example, the graph at the end of \rSC{sec:pet:termgraph} is a termination graph.
\rTH{Existence of
Termination Graphs} shows that termination graphs can always be obtained automatically.

\begin{theorem}[Existence of Termination Graphs]\label{Existence of
Termination Graphs}
For any program $\PP$ and abstract state $(S; \mathit{KB})$,
there exists a termination graph.
\end{theorem}

\begin{example}\label{ex:divminus:graphandlp}
For the
program from \rEX{ex:divminus}
we obtain the termination graph below.
Here, $\UU=
\{(\Fdiv(T_{5},T_{6},T_3),
\Fdiv(X,\Fz,Z)), (\Fdiv(T_{5},T_{6},T_3),\Fdiv(\Fz,Y,Z))\}$  results from exploiting the cuts.
$\UU$ implies that neither $T_6$ nor $T_5$ unify with $\Fz$. Thus,
only Clause (\ref{ex:divminus:6}) is applicable to evaluate the state in Node \textsc{d}. This is
crucial for termination, because in \textsc{d}, $\Fminus$'s result $T_{8}$ is always
smaller than $\Fminus$'s input argument $T_5$ and therefore,  $\Fdiv$'s first argument in
Node \textsc{c} is
smaller than $\Fdiv$'s first argument in Node \textsc{a}.

Remember \pagebreak that our goal is to show termination of the graph's
initial state. Since  the graph only has leaves $(\varepsilon;\mathit{KB})$
that are 
clearly terminating,
by soundness of the\linebreak inference rules, it remains to
prove that there is no state-derivation corresponding to an infinite
traversal of the
cycles in the graph. So in our example, we have to
show that the \textsc{Instance} edges for $\Fdiv$ and $\Fminus$ cannot be
traversed infinitely often.
\end{example}

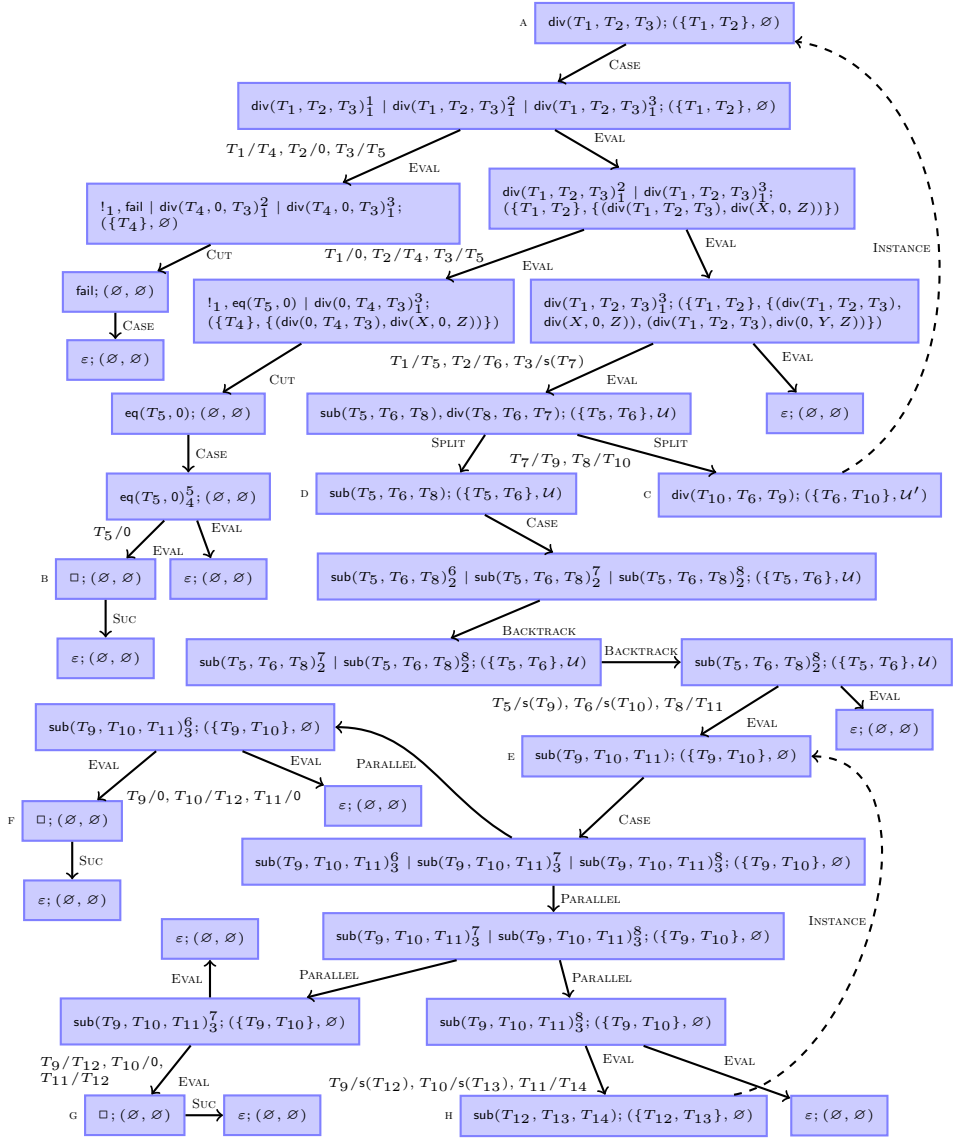
\input{divminusgraph}

We now synthesize a cut-free program from the termination
graph. This program has the following property: if there is a state-derivation
from a concretization of one state to a
concretization of another state which may be crucial for non-termination,
then there is a corresponding derivation in the
obtained cut-free program.

More precisely, we build clauses for all \emph{clause paths}.
For a termination graph $G$,
let \textsc{Instance}$(G)$ denote all nodes of $G$ to which the rule
\textsc{Instance} has been applied 
(i.e., \textsc{c} and \textsc{h} in our example).  The sets
\textsc{Split}$(G)$ and \textsc{Suc}$(G)$ are defined analogously.
For any node $n$, let  $Succ(i,n)$ denote the $i$-th child of $n$. Clause paths
are paths in the graph that start in the root node, in
the successor node of an \textsc{Instance} node, or in
the left child of a
\textsc{Split} node and
that end in a \textsc{Suc} or \textsc{Instance} node or
in the left child of an  \textsc{Instance} or
\textsc{Split} node.

\begin{definition}[Clause Path]
A path $\pi = n_{1} \ldots n_{k}$ in  $G$
is a \emph{clause path} iff $k > 1$ and
\vspace*{-.1cm}
\begin{itemize}
\item $n_{1} \in Succ(1,\textsc{Instance}(G) \cup \textsc{Split}(G))$ or $n_1$
is the root of $G$,
\item $n_{k} \in \textsc{Suc}(G) \cup \textsc{Instance}(G)  \cup Succ(1,\textsc{Instance}(G) \cup \textsc{Split}(G))$,
\item for all $1 \leq j < k$, we have $n_{j} \not\in \textsc{Instance}(G)$, and
\item for all $1 < j < k$, we have $n_{j} \not\in Succ(1,\textsc{Instance}(G) \cup \textsc{Split}(G))$.
\end{itemize}
\end{definition}

Since we only want finitely many clause paths, they  may
not traverse \textsc{Instance} edges. Clause paths may also not
follow left successors of \textsc{Instance} or \textsc{Split}. Instead, we create
new  clause paths starting at
these nodes.
In our example, we have clause paths from \textsc{a} to \textsc{b}, \textsc{a} to \textsc{c}, \textsc{a} to \textsc{d}, \textsc{d}
to \textsc{e}, \textsc{e} to \textsc{f}, \textsc{e} to \textsc{g}, and \textsc{e} to \textsc{h}.

To obtain a cut-free logic program, we construct one clause for each clause path $\pi = n_{1} \ldots n_{k}$.
The head of the new clause corresponds to $n_{1}$ where we apply the relevant
substitutions between $n_{1}$ and $n_{k}$. The last body atom corresponds to
$n_{k}$. The intermediate body atoms correspond to those nodes
that are left children of those $n_{i}$ which are from $\textsc{Split}(G)$.
Note that
we apply the relevant substitutions between $n_{i}$ and $n_{k}$ to the respective
intermediate body atom as well.

In our example,
the path from \textsc{a} to \textsc{b} is labeled by the substitution $\sigma =
\{ T_{1}/\Fz,\linebreak T_2/T_4, \,
T_{3}/\Fz, \, T_{5}/\Fz\}$. Hence, we
obtain the fact $\Fdiv_{\textsc{a}}(T_{1},T_{2},T_{3})\sigma = \Fdiv_{\textsc{a}}(\Fz,T_{4},\Fz)$.
We always use a new predicate symbol when translating a
node into an atom of a new clause (i.e., $\Fdiv_{\textsc{a}}$ is fresh).
\textsc{Instance} nodes are the only exception. There, we use the same predicate
symbol both for the \textsc{Instance} node and its successor.

For the path from \textsc{a} to \textsc{c},  we have  the
substitution  $\sigma' = \{T_{1}/T_{5}, \, T_{2}/T_{6}, \, T_3/\Fs(T_{9}),\linebreak
T_7/T_9, \, T_8/T_{10}\}$. Right children of \textsc{Split} nodes can only be reached if the goal in the
left \textsc{Split}-child  was successful.
So  $\Fminus(T_{5},T_{6},T_8)\sigma'$ must be derived to $\Box$
before the derivation can continue with $\Fdiv$.
Thus, we obtain the new clause $\Fdiv_{\textsc{a}}(T_5, T_6, \Fs(T_9)) \from
\Fminus_{\textsc{d}}(T_5,T_6,T_{10}), \Fdiv_{\textsc{a}}(T_{10},T_6,T_9)$. Note that we used the same
symbol $\Fdiv_{\textsc{a}}$ for both occurrences of $\Fdiv$ as they are
linked by an \textsc{Instance} edge.

Continuing in this
way, we obtain the following logic program for which we have to
show termination w.r.t.\ the set of
queries $\{\Fdiv_{\textsc{a}}(t_1,t_2,t_3)
 \mid t_1, t_2\text{ are ground}\}$, as
specified by the knowledge base in the root node \textsc{a}.
\addtocounter{equation}{1}
\setcounter{dmd}{\value{equation}}
\[ \begin{array}{rcl@{\qquad}l}
 \Fdiv_{\textsc{a}}(\Fz,T_4,\Fz).&&&\\
\Fdiv_{\textsc{a}}(T_5, T_6, \Fs(T_9)) &\from& \Fminus_{\textsc{d}}(T_5,T_6,T_{10}),
\Fdiv_{\textsc{a}}(T_{10},T_6,T_9). & (\arabic{dmd})\\
\Fdiv_{\textsc{a}}(T_5, T_6, \Fs(T_7)) &\from& \Fminus_{\textsc{d}}(T_5,T_6,T_8).&\\
\Fminus_{\textsc{d}}(\Fs(T_9),\Fs(T_{10}),T_{11})& \from& \Fminus_{\textsc{e}}(T_9,T_{10},T_{11}).&\\
\Fminus_{\textsc{e}}(\Fz,T_{12},\Fz).&&&\\
\Fminus_{\textsc{e}}(T_{12},\Fz,T_{12}).&&&\\
\Fminus_{\textsc{e}}(\Fs(T_{12}),\Fs(T_{13}),T_{14}) &\from& \Fminus_{\textsc{e}}(T_{12}, T_{13}, T_{14}).&
\end{array}\]

Virtually all existing methods and tools
for proving termination of logic programs succeed
on this definite logic program. Hence, by our \pagebreak pre-processing technique,
termination of programs with cut like \rEX{ex:divminus} can be proved automatically.

In general, to convert a node $n$ into an atom, we use a function $Ren$. $Ren(n)$ has the form
$p_n(X_1,\ldots,X_n)$ where $p_n$ is a fresh predicate symbol for the node $n$
(except if $n$ is an
\textsc{Instance} node) and $X_1,\ldots,X_n$ are all variables in $n$. This renaming allows us to use
different predicate symbols for different nodes. For example,
the cut-free logic program above would not terminate if we identified
$\Fminus_{\textsc{d}}$ and $\Fminus_{\textsc{e}}$. The reason is that $\Fminus_{\textsc{d}}$ only succeeds if its first
and second argument start with  ``$\Fs$''. Hence, if the intermediate body atom
$\Fminus_{\textsc{d}}(T_5,T_6,T_{10})$ of Clause (\arabic{dmd}) succeeds, then the ``number $T_{10}$'' will always be
strictly smaller than the ``number $T_5$''.
Finally, $Ren$ allows us to represent a whole state by just
one atom, even if this state consists of a non-atomic goal or a backtracking sequence with several
elements.

The only remaining problem is that paths may contain
evaluations for several alternative backtracking goals of the same case analysis.
Substitutions that correspond to ``earlier''
alternatives
must not be
regarded when instantiating the head of the new clause.
The reason is that backtracking
undoes the substitutions of previous evaluations. Thus, we collect the
substitutions on the path starting with the substitution
applied last. Here, we always keep track of the mark $d$ corresponding to the last
\textsc{Eval} node.
Substitutions that belong to earlier alternatives of
the
current case analysis are
disregarded when constructing the new cut-free program.
These earlier alternatives
can be identified easily, since they have marks  $m$ with $m \geq	d$.
 \begin{wrapfigure}[21]{r}{0.28\textwidth}
\vspace*{-.5cm}
\hspace*{-.2cm}
{\tiny
\begin{tikzpicture}
[node/.style={rectangle,draw=blue!50,fill=blue!20,thick,inner sep=5pt},
pre/.style={<-,thick}]
\begin{scope}[node distance=0.75 and 0.75]
\node[node,label=180:\textsc{a}] (bt1) {$\Fp(T_1)$};
\node[node] (bt2)  [below=.5 of bt1] {$\Fp(T_1)_1^{\tiny \ref{prob1}} \mid
\Fp(T_1)_1^{\tiny \ref{prob2}}$}
  edge [pre] node[auto,swap] {\textsc{Case}} (bt1);
\node[node] (bt3)  [below=.5of bt2] {$\Fq(T_2) \mid
\Fp(T_1)_1^{\tiny \ref{prob2}}$}
  edge [pre] node[auto,swap] {\textsc{Eval}}   node[auto] {$T_1/\Ff(T_2)$}
 (bt2);
\node[node] (bt4)  [below right =.7of bt2,xshift=-.4cm] {$\Fp(T_1)_1^{\tiny \ref{prob2}}$}
  edge [pre] node[auto,swap] {\textsc{Eval}}  (bt2);
\node[inner sep=5pt] (btdots)  [below=.5of bt4] {$\ldots$}
  edge [pre] node[auto,swap] {} (bt4);
\node[node] (bt3a)  [below=.5of bt3] {$\Fq(T_2)_2^{\tiny \ref{prob3}} \mid
\Fp(T_1)_1^{\tiny \ref{prob2}}$}
  edge [pre] node[auto,swap] {\textsc{Case}}
 (bt3);
\node[node,label=180:\textsc{b}] (bt3b)  [below=.5of bt3a] {$\Box \mid
\Fp(T_1)_1^{\tiny \ref{prob2}}$}
  edge [pre] node[auto,swap] {\textsc{Eval}}   node[auto] {$T_2/\Fa$}
 (bt3a);
\node[node] (bt3c)  [below right =.7of bt3a,xshift=-.6cm] {$\Fp(T_1)_1^{\tiny \ref{prob2}}$}
  edge [pre] node[auto,swap] {\textsc{Eval}}  (bt3a);
\node[inner sep=5pt] (btdots)  [below=.5of bt3c] {$\ldots$}
  edge [pre] node[auto,swap] {} (bt3c);
\node[node] (bt5)  [below =.5 of bt3b] {$\Fp(T_1)_1^{\tiny \ref{prob2}}$}
  edge [pre] node[auto,swap] {\textsc{Suc}}   (bt3b);
\node[node] (bt6)  [below =.5of bt5] {$\Fr(T_3)$}
  edge [pre] node[auto,swap] {\textsc{Eval}}    node[auto] {$T_1/\Fg(T_3)$}
 (bt5);
\node[node] (bt7)  [below right=.7of bt5] {$\varepsilon$}
  edge [pre] node[auto,swap] {\textsc{Eval}}  (bt5);
\node[node] (bt8)  [below =.5 of bt6] {$\Fr(T_3)_3^{\tiny \ref{prob4}}$}
  edge [pre] node[auto,swap] {\textsc{Case}}    (bt6);
\node[node,label=180:\textsc{c}] (bt9)  [below =.5of bt8] {$\Box$}
  edge [pre] node[auto,swap] {\textsc{Eval}}     node[auto] {$T_3/\Fb$}
 (bt8);
\node[node] (bt10)  [below right=.7of bt8] {$\varepsilon$}
  edge [pre] node[auto,swap] {\textsc{Eval}}    (bt8);
\node[node] (bt11)  [below =.5 of bt9] {$\varepsilon$}
  edge [pre] node[auto,swap] {\textsc{Suc}}     (bt9);
\end{scope}
\end{tikzpicture}}
\end{wrapfigure}

\addtocounter{example}{1}
\setcounter{problem ex}{\value{example}}

\noindent
\textit{Example \arabic{ex:pq}}

\noindent
Consider the
following  program and
 the termination graph  for the state  $(\Fp(T_1);
(\varnothing,\varnothing))$ on the side. Here,
we omitted the knowledge bases
to ease readability.

\vspace*{-.1cm}

\noindent
\hspace*{-.5cm}\begin{minipage}{5cm}
\begin{eqnarray}
\label{prob1}
\Fp(\Ff(X)) &\!\!\from\!\!& \Fq(X).\\
\label{prob2}
\Fp(\Fg(X)) &\!\!\from\!\!& \Fr(X).
\end{eqnarray}
\end{minipage}
\begin{minipage}{3cm}
\begin{eqnarray}
\label{prob3}
&&\Fq(\Fa).\\
\label{prob4}
&&\Fr(\Fb).
\end{eqnarray}
\end{minipage}

\vspace*{.2cm}

This graph contains clause paths from \textsc{a} to \textsc{b} and from \textsc{a} to
\textsc{c}.  For every clause path, we
collect the relevant substitutions step by step, starting from the end
of the path.
So for the first clause path 
we start with $\{T_2 / \Fa\}$. This substitution results from
an  \textsc{Eval} node for the goal
$\Fq(T_2)_2^{\footnotesize \ref{prob3}}$
 with mark $d = 2$. Hence,  for the first clause
path
we only collect further substitutions that
result from \textsc{Eval} nodes with marks smaller than $d = 2$. Since the next substitution
$\{ T_1 / \Ff(T_2)\}$ results
from an \textsc{Eval} node with mark 1, we finally obtain  $\{ T_1 / \Ff(T_2)\} \circ
\{T_2 / \Fa\}$
which leads to the  fact $\Fp(\Ff(\Fa))$ in the resulting logic
program.
For the second clause path from \textsc{a} to \textsc{c},
we start with $\{ T_3/\Fb \}$ which results from an
\textsc{Eval} node with mark $d = 3$. When moving upwards in the tree, the substitution $\{ T_1 /
\Fg(T_3) \}$ also has to be collected, since it results from an \textsc{Eval} node with
mark 1. Thus, we now set $d = 1$. When moving upwards, we reach further
substitutions, but they result from \textsc{Eval} nodes with marks 2 and 1.
These substitutions are not collected, since they correspond to
earlier alternatives of this case analysis.
Hence, we just obtain the substitution $\{ T_1 /
\Fg(T_3) \} \circ \{ T_3/\Fb \}$ for the second clause path,
which yields the fact $\Fp(\Fg(\Fb))$ in the resulting logic
program.

If we disregarded the marks when collecting substitutions, the second clause path
would result in  $\{ T_1 / \Ff(T_2)\} \circ
\{T_2 / \Fa\} \circ \{ T_1 /
\Fg(T_3) \} \circ \{ T_3/\Fb \}$ instead. But then we would get the same fact
$\Fp(\Ff(\Fa))$ as from
the first clause path. So the new logic program
would not simulate all derivations represented in the termination graph.

Now we formally define the cut-free logic program $\PP_G$ and the corresponding class of
queries $\QQ_G$ resulting from a termination graph $G$. If $\PP_G$ is terminating for all  queries from $\QQ_G$,
then the root state of $G$ is
terminating w.r.t.\ the original logic program (possibly containing
cuts).

\begin{definition}[Logic Programs and Queries from Termination Graph]\label{Logic Programs and Queries from Termination Graph}
Let $G$ be a termination graph whose root $n$ is
{\small $(p(T_1,...,T_m),
(\{T_{i_1},...,T_{i_k}\},\varnothing))$.}
 We define $\PP_{G} = \bigcup_{\pi\text{ clause
path in }G} \; Clause(\pi)$
and $\QQ_G= \{
 p_n(t_1,...,t_m) \mid t_{i_1},..., t_{i_k}$\linebreak
are ground$\}$. Here, $p_n$ is a new predicate which results from
translating the node $n$ into a clause.
For a path $\pi = n_{1} ... n_{k}$, let $Clause(\pi) = Ren(n_{1})\sigma_{\pi,\infty} \from
I_{\pi}, Ren(n_{k})$. For $n \in \textsc{Suc}(G)$, $Ren(n)$ is $\Box$ and for $n \in
\textsc{Instance}(G)$, it is $Ren(Succ(1,n))\mu$ where $\mu$ is the substitution associated with
the \textsc{Instance} node $n$. Otherwise, $Ren(n)$ is $p_{n}(\VV(n))$ where $p_{n}$ is a fresh
predicate symbol and $\VV(S;\mathit{KB}) = \VV(S)$.

Finally, $\sigma_{\pi,d}$ with $d \in \N \cup \{ \infty \}$ and $I_{\pi}$ are defined as
follows. Here for  a path $\pi = n_{1}
\ldots n_{j}$, the substitutions $\mu$ and $\sigma$ are the labels on the outgoing edge
of $n_{j-1} \in \textsc{Split}(G)$ and $n_{j-1} \in \textsc{Eval}(G)$, respectively, and the mark $m$
results from the corresponding node $n_{j-1} = ((t, Q)^i_m | S ; \mathit{KB})$.

\vspace*{-.3cm}

{\small
\[\sigma_{n_{1} \ldots n_{j}, d} = \begin{cases}
id & \text{if $j = 1$}\\
\sigma_{n_{1} \ldots n_{j-1},d} \; \mu & \text{if $n_{j-1} \in \textsc{Split}(G)$, $n_{j} = Succ(2,n_{j-1})$}\\
\sigma_{n_{1} \ldots n_{j-1},m} \; \sigma & \text{if $n_{j-1} \in \textsc{Eval}(G)$,
$n_{j} = Succ(1,n_{j-1})$, and $d > m$}\\
\sigma_{n_{1} \ldots n_{j-1},d} \; \sigma\mbox{\scriptsize $|$}_{\GG} & \text{if $n_{j-1}
\in \textsc{Eval}(G)$, $n_{j} = Succ(1,n_{j-1})$, and $d \leq m$}\\
\sigma_{n_{1} \ldots n_{j-1},d} & \text{otherwise}
\end{cases}\]
\vspace*{-.3cm}
\[I_{n_{j} \ldots n_{k}} = \begin{cases}
\Box & \text{if $j = k$}\\
Ren(Succ(1,n_{j}))\sigma_{n_{j}\ldots n_{k},\infty}, \; I_{n_{j+1} \ldots n_{k}} & \text{if $n_{j} \in \textsc{Split}(G), n_{j+1} = Succ(2,n_{j})$}\\
I_{n_{j+1} \ldots n_{k}} & \text{otherwise}
\end{cases}\]}
\end{definition}

So if $n_{j-1}$ is a \textsc{Split} node, then one has to ``collect'' the corresponding
substitution $\mu$ when constructing the overall substitution $\sigma_{n_1 \ldots n_j, d}$
for the path. If $n_{j-1}$ is an \textsc{Eval} node for the $m$-th case analysis
and $n_j$ is its left successor,
then the construction of $\sigma_{n_1 \ldots n_j, d}$ depends on whether we have
already collected a corresponding substitution for the current case analysis $m$.
If $m$ is smaller than the mark $d$ for the last case analysis which contributed
to the substitution, then the corresponding substitution $\sigma$ of the
\textsc{Eval} rule is collected and $d$ is set to $m$. Otherwise (if $d \leq m$), one only collects the
part $\sigma\mbox{\scriptsize $|$}_{\GG}$ of the
substitution that concerns those abstract variables that stand for ground terms.
The definition of the intermediate body atoms $I_\pi$ ensures that 
derivations in $\PP_G$ only reach the second child of a \textsc{Split} node if the first child of
the \textsc{Split} node could successfully be proved.

\rTH{thm:pet:correctness} proves the soundness of our approach. So termination of the
cut-free program $\PP_G$ implies termination of the original
program $\PP$. \pagebreak (However as shown in \cite{techreport}, the converse does not hold.)

\begin{theorem}[Soundness]\label{thm:pet:correctness}
Let $G$ be a proper termination graph for  $\PP$
whose root  is
{\small $(p(T_1,...,T_m),
(\{T_{i_1},...,T_{i_k}\},$}\linebreak
{\small $\varnothing))$.}
If $\PP_G$ terminates for all queries in $\QQ_G$,
then all concretizations
of  $G$'s root state have only finite state-derivations. In other words,
then all queries from the set $\{ p(t_1,\ldots,t_m) \mid  t_{i_1},\ldots,t_{i_k} \mbox{ are ground}\}$
terminate w.r.t.\
$\PP$.
\end{theorem}

\section{Experiments and Conclusions}\label{sec:pet:conclusion}

We introduced a pre-processing method to eliminate
cuts. Afterwards, any technique for proving
universal termination of logic programming can be applied.
Thus, termination of logic programs with cuts can now be
analyzed automatically.

We implemented this pre-processing  in our tool \aprove{} \cite{APROVE:IJCAR06}
and performed extensive experiments which show that now we can indeed prove
termination of typical logic programs with cut fully automatically. The
implementation is not only successful for programs like \rEX{ex:divminus}, but also for
programs using operators\linebreak like
 \emph{negation as failure}
or \emph{if then else}
which can be expressed using cuts.
While \aprove{} was already very
powerful for termination analysis of definite logic programs
\cite{TOCL08}, our pre-processing method strictly increa\-ses its
power.
For our experiments, we used the \emph{Termination Problem Database} (TPDB)
of the annual \emph{International Termination
Competition}.\footnote{\url{http://termination-portal.org/wiki/Termination_Competition}}
Since up to now, no tool had special support for cuts, the previous versions
of the TPDB did not contain any programs with cuts. Therefore, we took
existing cut-free examples from the TPDB and added cuts in a natural way. In
this way, we extended the TPDB by 104 typical programs with cuts
(directory {\tt LP/CUT}).
Of these,
10  are known to be non-terminating.
Up to now, termination tools treated cuts by simply ignoring them and by
trying to prove termination of the program that results from removing the
cuts. This is sensible, since cuts are not always needed for
termination. Indeed, a
version of \aprove\ that ignores
cuts and does not use our pre-processing can show termination of 10 of the 94 potentially
terminating examples.
Other existing termination tools
would not yield much better results, since \aprove\ is already the most powerful tool
for definite logic programming (as shown by the experiments in \cite{TOCL08})
and since  most of the
remaining 84 examples do not terminate anymore if one removes the
cut.
In contrast, with our new pre-processing,
\aprove\ proves termination of 78 examples (i.e., 83\% of the
potentially terminating examples). This  shows that our contributions
are crucial for termination analysis of logic programs with cuts.
Nevertheless, there is of course  room for further improvements (e.g.,
one could develop alternative 
techniques to generate cut-free clauses from the termination graph
in order
to improve the performance on examples which encode existential termination).
To experiment with our implementation and for further details, we refer to 
\url{http://aprove.informatik.rwth-aachen.de/eval/Cut/}.

\vspace{.3cm}

\noindent
\textbf{Acknowledgements.} We thank the referees for many helpful remarks.

\bibliography{references}

\end{document}

%% file: commands.tex
\newcounter{mytmpEquation}
\newcounter{tmpFooter}

\newcounter{tmp}
\newcounter{dpcounter}
\newcommand{\introproof}[3]{%
\expandafter\def\csname proof#2\endcsname{%
\begin{proof}[Proof of #1 \ref{#2}]
\label{proofof#2}
 #3%
\end{proof}%
}%
}

\newcommand{\showproof}[1]{%
\csname proof#1\endcsname}

\newcommand{\ito}[1]%
{\mathrel{\smash{\stackrel{\raisebox{2pt}{\scriptsize $\mathsf{i}\:$}}%
{\smash{\rightarrow}}}_{#1}}}
\newcommand{\itos}[2]%
{\mathrel{\smash{\stackrel{\raisebox{2pt}{\scriptsize $\mathsf{i}\:$}}%
{\smash{\rightarrow}}}_{#1}^{#2}}}

\newcommand{\hideproof}[1]{}

\renewcommand{\emptyset}{\varnothing}
\newcommand{\aprove}{{\sf AProVE}}
\newcommand{\rDF}[1]{\text{Def.\ \ref{#1}}}

\newcommand{\rEX}[1]{\text{Ex.\ \ref{#1}}}

\newcommand{\rTH}[1]{\text{Thm.\ \ref{#1}}}
\newcommand{\rSC}[1]{Sect.\ \ref{#1}}

\newcommand{\GG}{\mathcal{G}}
\newcommand{\NN}{\mathcal{N}}

\newcommand{\VV}{\mathcal{V}}
\newcommand{\TT}{\mathcal{T}}
\newcommand{\UU}{\mathcal{U}}
\newcommand{\PP}{\mathcal{P}}

\newcommand{\SSS}{\mathcal{C}on}
\newcommand{\QQ}{\mathcal{Q}}

\newcommand{\Ff}{\mathsf{f}}

\newcommand{\Fg}{\mathsf{g}}

\newcommand{\Fa}{\mathsf{a}}

\newcommand{\Fb}{\mathsf{b}}

\newcommand{\Fr}{\mathsf{r}}
\newcommand{\Fs}{\mathsf{s}}

\newcommand{\Fz}{\mathsf{0}}

\newcommand{\Fp}{\mathsf{p}}

\newcommand{\Fq}{\mathsf{q}}
\newcommand{\Fdiv}{\mathsf{div}}

\newcommand{\Fminus}{\mathsf{sub}}

\renewcommand{\epsilon}{\varepsilon}

\newcommand{\Slice}{\mathit{Slice}}

\newcommand{\iindex}[1]{}

\newcommand{\newwindex}[1]{%
\index{\csname #1idx\endcsname}}

\newcommand{\procindex}[1]{%
\glossary{#1}}

\newcommand{\introProc}[1]{%
\procindex{\csname name#1\endcsname}%
\label{Processor #1}%
}

\newcommand{\nameProc}[1]{%
\csname name#1\endcsname}

\newcommand{\eat}[1]{}

\newcommand{\N}{\mathbb{N}}
\newcommand{\from}{\ensuremath{\leftarrow}}

\renewcommand{\AA}{\mathcal{A}}

\newcommand{\customeqn}[6]{\setcounter{saveeqn}{\value{equation}}\setcounter{equation}{0}\renewcommand{\theequation}{%
  \ifthenelse{\equal{\value{equation}}{1}}{#1}{}%
  \ifthenelse{\equal{\value{equation}}{2}}{#2}{}%
  \ifthenelse{\equal{\value{equation}}{3}}{#3}{}%
  \ifthenelse{\equal{\value{equation}}{4}}{#4}{}%
  \ifthenelse{\equal{\value{equation}}{5}}{#5}{}%
  \ifthenelse{\equal{\value{equation}}{6}}{#6}{}%
}}
\newcommand{\reseteqn}{\renewcommand{\theequation}{\arabic{equation}}\setcounter{equation}{\value{equation}}}

\renewcommand{\P}{\mathcal{P}}
\renewcommand{\S}{\mathcal{S}}

\newcommand{\State}{\mathit{State}}
\newcommand{\Goal}{\mathit{Goal}}
\newcommand{\AState}{\mathit{AState}}

\newcommand{\Ffail}{{\sf fail}}
\newcommand{\Ffailure}{{\sf failure}}

%% file: divminusgraph.tex
{\tiny
\begin{center}
\begin{tikzpicture}
[node/.style={rectangle,draw=blue!50,fill=blue!20,thick,inner sep=5pt},
pre/.style={<-,thick},post/.style={->,thick,dashed}]
\begin{scope}[node distance=0.65 and 0.65]
\node[node,label=180:\textsc{a}] (1) {$\Fdiv(T_1,T_2,T_3); (\{T_1, T_2\}, 
\emptyset)$};
\node[node] (2) [below=0.5 of 1,xshift=-20mm] {$\Fdiv(T_1,T_2,T_3)^{\ref{ex:divminus:1}}_1 \mid \Fdiv(T_1,T_2,T_3)^{\ref{ex:divminus:2}}_1 \mid \Fdiv(T_1,T_2,T_3)^{\ref{ex:divminus:3}}_1; (\{T_1, T_2\}, 
\emptyset)$}
  edge [pre] node[auto,swap,xshift=1.3mm,yshift=1.3mm] {\textsc{Case}} (1);
\node[node] (3) [below=.7 of 2,xshift=-3.2cm] {\begin{minipage}{4.6cm}$!_1, \Ffail \mid \Fdiv(T_4,\Fz,T_3)^{\ref{ex:divminus:2}}_1 \mid \Fdiv(T_4,\Fz,T_3)^{\ref{ex:divminus:3}}_1;\\(\{T_4\}, 
\emptyset)$\end{minipage}}
  edge [pre] node[auto,swap] {\textsc{Eval}} node[auto,xshift=-1mm,yshift=-1mm] {$T_1/T_4,T_2/\Fz,T_3/T_5$} (2);
\node[node] (4) [below=0.5 of 3,xshift=-2.1cm,yshift=.15cm] {$\Ffail; (\emptyset, 
\emptyset)$}
  edge [pre] node[auto,swap,yshift=.2cm,xshift=.2cm] {\textsc{Cut}} (3);
\node[node] (5') [below=0.5 of 4,yshift=.15cm] {$\varepsilon; (\emptyset, 
\emptyset)$}
  edge [pre] node[auto,swap] {\textsc{Case}} (4);
\node[node] (4') [below=.5 of 2,xshift=2.1cm] {\begin{minipage}{4.5cm}$\Fdiv(T_1,T_2,T_3)^{\ref{ex:divminus:2}}_1 \mid \Fdiv(T_1,T_2,T_3)^{\ref{ex:divminus:3}}_1 ;\\(\{T_1, T_2\},
\{(\Fdiv(T_1,T_2,T_3),\Fdiv(X,\Fz,Z))\})$\end{minipage}}
  edge [pre] node[auto,swap] {\textsc{Eval}} (2);
\node[node] (5) [below left=of 4',xshift=1cm] {\begin{minipage}{3.9cm}$!_1, {\sf eq}(T_5,\Fz) \mid \Fdiv(\Fz,T_4,T_3)^{\ref{ex:divminus:3}}_1;\\[0.5ex] (\{T_4\}, 
\{(\Fdiv(\Fz,T_4,T_3),\Fdiv(X,\Fz,Z))\})$\end{minipage}}
  edge [pre] node[auto,swap] {\textsc{Eval}} node[auto,xshift=-3mm,yshift=-2mm] {$T_1/\Fz,T_2/T_4,T_{3}/T_{5}$} (4');
\node[node] (6) [below=of 5,xshift=-2.2cm] {${\sf eq}(T_5,\Fz); (\emptyset, 
\emptyset)$}
  edge [pre] node[auto,swap] {\textsc{Cut}} (5);
\node[node] (66) [below=0.5 of 6] {${\sf eq}(T_5,\Fz)^{\ref{ex:divminus:7}}_{4}; (\emptyset, 
\emptyset)$}
  edge [pre] node[auto,swap] {\textsc{Case}} (6);
\node[node,label=180:\textsc{b}] (7) [below=0.5 of 66,xshift=-1.1cm] {$\Box; (\emptyset, 
\emptyset)$}
  edge [pre] node[auto,swap] {\textsc{Eval}} node[auto,xshift=-1mm,yshift=-1mm] {$T_5/\Fz$} (66);
\node[node] (8) [below=0.5 of 7] {$\varepsilon; (\emptyset, 
\emptyset)$}
  edge [pre] node[auto,swap] {\textsc{Suc}} (7);
\node[node] (7') [below=0.5 of 66,xshift=.4cm] {$\varepsilon; (\emptyset, 
\emptyset)$}
  edge [pre] node[auto,swap] {\textsc{Eval}} (66);
\node[node] (6'') [below=of 4',xshift=.7cm] {\begin{minipage}{4.8cm}$\Fdiv(T_1,T_2,T_3)^{\ref{ex:divminus:3}}_1; (\{T_1, T_2\}, 
 \{(\Fdiv(T_1,T_2,T_3),\\[0.5ex]\Fdiv(X,\Fz,Z)),(\Fdiv(T_1,T_2,T_3),\Fdiv(\Fz,Y,Z))\})$\end{minipage}}
  edge [pre] node[auto,swap] {\textsc{Eval}} (4');
\node[node] (8) [below=of 6'',xshift=-3cm] {$\Fminus(T_{5},T_{6},T_8), \Fdiv(T_8,T_{6},T_{7}); (\{T_{5}, T_{6}\}, \UU)$}
  edge [pre] node[auto,swap] {\textsc{Eval}} node[auto,xshift=-1mm,yshift=-1mm] {$T_{1}/T_{5}, T_{2}/T_{6}, T_{3}/\Fs(T_{7})$} (6'');
\node[node] (7'') [below=of 6'',xshift=1.2cm] {$\varepsilon; (\emptyset,  \emptyset)$}
  edge [pre] node[auto,swap] {\textsc{Eval}} (6'');
\node[node,label=180:\textsc{d}] (9) [below=0.5of 8,xshift=-.7cm] {$\Fminus(T_{5},T_{6},T_8); (\{T_{5}, T_{6}\}, \UU)$}
  edge [pre] node[auto] {\textsc{Split}} (8);
\node[node,label=180:\textsc{c}] (9') [below=0.5of 8,xshift=4cm] {$\Fdiv(T_{10},T_{6},T_{9}); (\{T_{6}, T_{10}\},  \UU')$}
  edge [pre] node[auto,swap] {\textsc{Split}} node[auto,xshift=-1mm,yshift=1mm] {$T_7/T_9,T_8/T_{10}$} (8)
  edge [post,out=28,in=-8] node[auto] {\textsc{Instance}} (1);
\node[node] (10) [below=.5of 9,xshift=2cm] {$\Fminus(T_{5},T_{6},T_{8})^{\ref{ex:divminus:4}}_2 \mid \Fminus(T_{5},T_{6},T_{8})^{\ref{ex:divminus:5}}_2 \mid \Fminus(T_{5},T_{6},T_{8})^{\ref{ex:divminus:6}}_2; (\{T_{5}, T_{6}\}, \UU)$}
  edge [pre] node[auto,swap] {\textsc{Case}} (9);
\node[node] (11') [below =.5of 10,xshift=-2.7cm] {$\Fminus(T_{5},T_{6},T_8)^{\ref{ex:divminus:5}}_2 \mid \Fminus(T_{5},T_{6},T_8)^{\ref{ex:divminus:6}}_2; (\{T_{5}, T_{6}\}, \UU)$}
  edge [pre] node[auto,swap] {\textsc{Backtrack}} (10);
\node[node] (12'') [right=of 11',xshift=.4cm] {$\Fminus(T_{5},T_{6},T_8)^{\ref{ex:divminus:6}}_2; (\{T_{5}, T_{6}\}, \UU)$}
  edge [pre] node[auto,swap] {\textsc{Backtrack}} (11');
\node[node,label=180:\textsc{e}] (13') [below=of 12'',xshift=-2cm] {$\Fminus(T_{9},T_{10},T_{11}); (\{T_{9}, T_{10}\}, 
\emptyset)$}
  edge [pre] node[auto,swap] {\textsc{Eval}} node[auto,xshift=-1mm,yshift=-1mm] {$T_{5}/\Fs(T_{9}), T_{6}/\Fs(T_{10}), T_8/T_{11}$} (12'');
\node[node] (13'') [below=.3 of 12'',xshift=.9cm] {$\varepsilon; (\emptyset, 
\emptyset)$}
  edge [pre] node[auto,swap,yshift=-1.3mm,xshift=1.3mm] {\textsc{Eval}} (12'');
\node[node] (14) [below=.8 of 13',xshift=-1.5cm] {$\Fminus(T_{9},T_{10},T_{11})^{\ref{ex:divminus:4}}_3 \mid \Fminus(T_{9},T_{10},T_{11})^{\ref{ex:divminus:5}}_3 \mid \Fminus(T_{9},T_{10},T_{11})^{\ref{ex:divminus:6}}_3; (\{T_{9}, T_{10}\}, 
\emptyset)$}
  edge [pre] node[auto,swap] {\textsc{Case}} (13');
\node[node] (15) [above left=of 14,xshift=1.9cm,yshift=.5cm] {$\Fminus(T_{9},T_{10},T_{11})^{\ref{ex:divminus:4}}_3; (\{T_{9}, T_{10}\}, 
\emptyset)$}
  edge [pre,in=150,out=0] node[auto,swap,yshift=2mm,xshift=-0.5mm] {\textsc{Parallel}} (14);
\node[node] (15') [below=.35 of 14] {$\Fminus(T_{9},T_{10},T_{11})^{\ref{ex:divminus:5}}_3 \mid \Fminus(T_{9},T_{10},T_{11})^{\ref{ex:divminus:6}}_3; (\{T_{9}, T_{10}\}, 
\emptyset)$}
  edge [pre] node[auto,swap] {\textsc{Parallel}} (14);
\node[node,label=180:\textsc{f}] (16) [below=of 15,xshift=-1.5cm] {$\Box; (\emptyset, 
\emptyset)$}
  edge  [pre] node[auto] {\textsc{Eval}}
node[auto,swap,xshift=-1mm,yshift=-1mm] {$T_{9}/\Fz, T_{10}/T_{12}, T_{11}/\Fz$} (15);
\node[node] (16') [below=.45 of 15,xshift=2.5cm] {$\varepsilon; (\emptyset, 
\emptyset)$}
  edge [pre] node[auto,swap,yshift=-0.5mm] {\textsc{Eval}} (15);
\node[node] (16'') [below left=.7 of 15',xshift=1cm] {$\Fminus(T_{9},T_{10},T_{11})^{\ref{ex:divminus:5}}_3; (\{T_{9}, T_{10}\}, 
\emptyset)$}
  edge [pre] node[auto,yshift=-1mm,xshift=-2mm] {\textsc{Parallel}} (15');
\node[node] (16''') [below=.5 of 15',xshift=.3cm] {$\Fminus(T_{9},T_{10},T_{11})^{\ref{ex:divminus:6}}_3; (\{T_{9}, T_{10}\}, 
\emptyset)$}
  edge [pre]  node[auto,swap,yshift=-1.3mm] {\textsc{Parallel}} (15');
\node[node] (17) [below=.5 of 16] {$\varepsilon; (\emptyset, 
\emptyset)$}
  edge [pre] node[auto,swap] {\textsc{Suc}} (16);
\node[node,label=180:\textsc{g}] (17') [below=of 16'',xshift=-1cm] {$\Box; (\emptyset, 
\emptyset)$}
  edge [pre] node[auto,swap] {\textsc{Eval}} node[auto,xshift=-1mm,yshift=-3mm] 
{\begin{minipage}{15.5mm}$T_{9}/T_{12},T_{10}/\Fz,\linebreak T_{11}/T_{12}$\end{minipage}} (16'');
\node[node] (17'') [above=.5 of 16''] {$\varepsilon; (\emptyset, 
\emptyset)$}
  edge [pre] node[auto,swap] {\textsc{Eval}} (16'');
\node[node,label=180:\textsc{h}] (17''') [below=of 16''',xshift=.5cm] {$\Fminus(T_{12},T_{13},T_{14}); (\{T_{12}, T_{13}\}, \emptyset)$}
  edge [pre] node[auto,swap] {\textsc{Eval}} node[auto] {$T_{9}/\Fs(T_{12}), T_{10}/\Fs(T_{13}), T_{11}/T_{14}$} (16''')
  edge [post,out=10,in=0] node[auto,yshift=-2mm] {\textsc{Instance}} (13');
\node[node] (17'''') [below=of 16''',xshift=3.5cm] {$\varepsilon; (\emptyset, 
\emptyset)$}
  edge [pre] node[auto,swap] {\textsc{Eval}} (16''');
\node[node] (18) [right=.5 of 17'] {$\varepsilon; (\emptyset, 
\emptyset)$}
  edge [pre] node[auto,swap] {\textsc{Suc}} (17');
\end{scope}
\end{tikzpicture}
\end{center}
}